\documentstyle[prl,aps,preprint]{revtex}

\def\norc #1 {\ {\Vert} #1 {\Vert_{c}}}
\def\norN #1 {\ {\Vert} #1 {\Vert_{N}}}
\def\nor #1 {\ {\Vert} #1 {\Vert}}

\def \Ninf#1 {  {\Vert {#1} \Vert_\infty } }
\def \NLip#1 {  {\Vert {#1} \Vert_\theta } }
\def \Norm#1 {  {\Vert {#1} \Vert } }

\def \bra{{ \langle}}
\def \ket{{ \rangle}}

\def \lcIN { {\rm in} }

\def \col{ \mskip -2mu : \mskip -2mu}

\def \hik{ \mskip -2mu - \mskip -2mu}
\def \tas{ \mskip -2mu + \mskip -2mu}

\begin{document}
\draft
\title{An Analytical Construction of the SRB Measures \\ for Baker-type Maps}

\author{S. Tasaki}
\address{Department of Physics, Nara Women's University,\\
Nara 630, Japan, \\ and \\
Institute for Fundamental Chemistry, \\ 34-4 Takano Nishihiraki-cho,
Sakyo-ku, Kyoto 606, Japan}
\author{Thomas Gilbert and J. R. Dorfman}
\address{Institute for Physical Science and Technology, \\
and Department of Physics, \\
University of Maryland, \\ 
College Park, Maryland 20742-3511, U.S.A.}
\date{\today}
\maketitle

\newpage

\begin{abstract}

For a class of dynamical systems, called the axiom-A systems, Sinai, Ruelle and
Bowen showed the existence of an invariant measure (SRB measure) weakly
attracting the temporal average of any initial distribution that is absolutely
continuous with respect to the Lebesgue measure. Recently, the SRB measures were
found to be related to the nonequilibrium stationary state distribution 
functions for thermostated or open systems. Inspite of the importance of these
SRB measures, it is difficult to handle them analytically because they are
often singular functions. In this article, for 
three
kinds of
Baker-type maps, the SRB measures are analytically constructed with the aid of
a functional equation, which was proposed by de Rham in order to deal with a
class of  singular
functions. We first briefly review the properties of singular functions
including those of de Rham. Then, the Baker-type maps are described,
one of which is non-conservative but time reversible, the second has a
Cantor-like invariant set, and the third is a model of a simple chemical reaction
$R \leftrightarrow I \leftrightarrow P$.
For the second example, the cases with and without escape are considered.
For the last example, we consider the reaction processes in a closed system and
in an open system under a flux boundary condition.
In all cases, we show that the evolution equation of the distribution 
functions partially integrated over the unstable direction is very similar to de
Rham's functional equation and, employing this analogy, we explicitly construct
the SRB measures.

\end{abstract}


\newpage

{\bf

\centerline{LEAD PARAGRAPH}

\vskip 18 pt 

Characterization of nonequilibrium stationary states in terms of
dynamical ensembles is one of the main questions in statistical
mechanics. Recently, the so-called Sinai-Ruelle-Bowen (SRB)  measures,
which had been studied in dynamical systems theory, were found to be
related to nonequilibrium stationary ensembles for thermostated or open
systems. The SRB measures fully describe transport properties of the
corresponding nonequilibrium stationary states. Also they would provide
an important insight about the emergence of irreversibility in
reversible dynamical systems, since they do not have time-reversal
invariance even when the dynamics is reversible. It is therefore
illustrative to know exact forms of the SRB measures, but it is
difficult to handle them exactly because they are often singular
functions.  In this paper, we study three examples of Baker-type maps,
which illustrate some aspects of the thermostated and/or open systems:
One is non-conservative but time reversible, the second has a Cantor-like
invariant set, and the third is a model of a simple chemical reaction
such as an isomerization $R \leftrightarrow I \leftrightarrow P$.
For those maps, we analytically construct SRB measures with the aid of a
new method, where the weak convergence of measures is converted to the
strong convergence of partially integrated distribution functions
(PIDFs) and the evolution equations for the PIDFs are solved emplying
the analogy between them and de Rham's functional equations.

}

\newpage

\narrowtext

\section{STATISTICAL MECHANICS AND SRB MEASURES}


One of the main questions in statistical mechanics is to characterize
nonequilibrium stationary states in terms of dynamical ensembles (cf. e.g.,
Refs. \cite{1,KH,Rei}).  
Recently, for thermostated or open systems, stationary nonequilibrium ensembles
were found to be related to the so-called Sinai-Ruelle-Bowen (SRB) measures
\cite{6,7,8,9,10,11,12,13,14,15,16,17}, which had been
investigated in dynamical systems theory \cite{2,3,4,5}.
In the thermostated systems \cite{6,7,8,9,10,11,12,13,14}, a fictitious
damping force mimicking a heat reservoir
is introduced to avoid an uncontrolled growth of the kinetic energy due
to an external driving force. The damping force is chosen so as to make the
dynamics dissipative while  it preserves time-reversibility. 
As a result of the dissipation, there exists an attractor of 
information dimension less than the dimension of phase space
and the nonequilibrium stationary state is described by an
asymptotic measure, which is an SRB measure. 
The SRB measure fully characterizes the transport properties, such as the
transport law, transport coefficients and their fluctuations.
For example, for the driven thermostated Lorentz gas \cite{8}, the conductivity
tensor was calculated, and Ohm's law and Einstein's relation were 
verified by comparing the averaged current with respect to the SRB measure to
the external electric field. 
On the other hand, open chaotic Hamiltonian systems with a flux boundary
condition \cite{15,16,17} admit a nonequilibrium stationary state obeying
Fick's law that is described by a kind of SRB measure.
This measure again characterizes the transport properties. 
Moreover, an open Hamiltonian system with an absorbing boundary condition has a
fractal repeller, that controls the chaotic scattering
\cite{14,18,19,19b,20,20b,21,22}. The unstable manifold of the fractal repeller
supports a conditionally invariant measure, which provides the long time limits of
averaged dynamical functions \cite{20,20b}.
The interrelation between the thermostated systems approach and the open
systems approach has been discussed by  Breymann,  T\'el and Vollmer
\cite{23}.
In this article, we present an analytical construction of SRB measures for
three examples of Baker-type maps, which illustrate some aspects of the
thermostated and/or open systems mentioned above.
Now we start with the general arguments on the SRB measure.
  

The long-term behavior of a dynamical system is characterized by an invariant
measure $\mu$ on an invariant set $A$, which describes how frequently various
parts of $A$ are visited by a given orbit $x(t)$ (with $t$ the time).  The
invariant measure is said to be ergodic if it cannot be decomposed into
different invariant measures. 
Such an ergodic invariant measure $\mu$ satisfies the ergodic theorem
\cite{1,KH,Rei,2}.  In case of a map $S$, it asserts that, for any
continuous function $\varphi(x)$, we have

\begin{equation}
\lim_{T\to+\infty} {1\over T} \sum_{t=0}^T \varphi(S^tx) = {\displaystyle
\int_A \varphi(x')\mu(dx') \over \displaystyle  \int_A \mu(dx')} \ . 
\qquad ( x\in A\backslash E  \ {\rm with} \ \mu(E)=0)
\label{1.1}
\end{equation}

\noindent 
A dynamical system typically admits uncountably many distinct
ergodic measures and not all of them are physically observable.  One
criterion of
choosing a physical measure $\mu$ is that $\mu$ describes the time averages
of observables on motions with initial data $x$ randomly sampled with respect
to the Lebesgue measure $\mu_0$ \cite{2,3}:

\begin{equation}
\lim_{T\to+\infty} {1\over T} \sum_{t=0}^T \varphi(S^tx) = {\displaystyle
\int_A \varphi(x')\mu(dx') \over \displaystyle  \int_A \mu(dx')} \ . 
\qquad \left({\displaystyle x\in \Sigma\backslash E    \ {\rm with} \ \Sigma
\supset A
\ , \atop \displaystyle
\mu_0(\Sigma)>0 \  {\rm and}
\
\mu_0(E)=0}\right) \label{1.2}
\end{equation}

\noindent
Sinai, Ruelle and Bowen showed that a class of dynamical systems, called
axiom-A systems, uniquely admit such a physical invariant measure
(SRB measure) \cite{2,3,4}. Thus an SRB measure is one
for which the ergodic theorem is true for almost every point, $x$, with
respect to the Lebesgue measure $\mu_0.$

Axiom-A systems are characterized by the hyperbolic structure, i.e., the
existence of exponentially stable and exponentially unstable directions which
intersect transversally with each other. In case of bijective differentiable
maps (i.e., diffeomorphisms), the hyperbolic structure is defined as follows:
Consider a given orbit $S^tx$ ($t=0,\pm 1,\pm 2, \cdots$) and small deviations
$\delta x$ of the initial value $x$. Note that there are $m$ 
independent possible
directions of $\delta x$ when the phase space dimension is $m$. Assume that the
deviations along $m_s$ directions decrease more rapidly than an 
expontial function
$e^{-\lambda t}$ ($\lambda>0$ and  $t\ge 0$) and that the deviations along the
other  $m-m_s$ directions grow more rapidly than an exponential function
$e^{\lambda t}$ ($t\ge 0$ and the same $\lambda$). An invariant set $\Lambda$
is said to be hyperbolic when 1) \ for any $x\in \Lambda$, the orbit $S^tx$
has the
$m_s$ stable and $m-m_s$ unstable directions as explained above,  2) \ the
stable and unstable directions depend continuously on $x$ and 3) \ the stable
and unstable directions for the point $x$ are mapped by $S^t$ to the
corresponding directions for the point $S^tx$. 

A point $x$ is said to be nonwandering if the orbit $S^tx$ returns indefinitely
often to any neighborhood of its initial point $x$. If the set $\Omega$ of all
nonwandering points is hyperbolic and the set of periodic points is dense in 
$\Omega$, $S$ is called an axiom-A diffeomorphism. In particular, if the whole
phase space $M$ is hyperbolic, $S$ is called an Anosov diffeomorphism. The
Arnold cat map is an example of an Anosov diffeomorphism. 

Invariant measures that are smooth along the unstable directions are called
SRB measures. Sinai, Ruelle and Bowen showed that, for axiom-A diffeomorphisms,
the SRB measure is the unique physical measure $\mu$ describing the time
averages (\ref{1.2}) of observables of motion with initial data $x$ taken at
random with respect to the Lebesgue measure $\mu_0$ \cite{2,3,4}. For more
details on axiom-A systems and SRB measures, see Refs. \cite{2}, \cite{3} and
\cite{5}.

In Gibbs' picture of statistical mechanics, a macroscopic state for an isolated
system is described by a phase-space distribution function, and a macroscopic
observable by an averaged dynamical function with respect to the distribution.
Suppose that the dynamics satisfies the mixing condition with respect to the
microcanonical distribution $\mu_{mc}$:

\begin{equation}
\lim_{t\to+\infty} \int_M \varphi(S^tx) \rho_0(x) \mu_{mc}(dx) = {\displaystyle
\int_M \varphi(x)\mu_{mc}(dx) \over \displaystyle  \int_M \mu_{mc}(dx)} \ , 
\label{1.3}
\end{equation}

\noindent
where $M$ stands for the whole phase space, $\varphi(x)$ is a continuous 
dynamical function and $\rho_0(x)$ is a normalized initial distribution 
function.
Then, the system exhibits time evolution as expected from statistical 
thermodynamics. Namely,
the distribution function weakly approaches an equilibrium microcanonical
distribution and the averaged dynamical functions approach well defined 
equilibrium values
\cite{1}. From this point of view, instead of Eq. (\ref{1.2}), it is enough to
consider the following condition as a criterion of choosing a physical measure
$\mu$:

\begin{equation}
\lim_{t\to+\infty} \int_M \varphi(S^tx) \rho_0(x) \mu_0(dx) = {\displaystyle
\int_A \varphi(x)\mu(dx) \over \displaystyle  \int_A \mu(dx)} \ , 
\label{1.4}
\end{equation}

\noindent
where $\mu_0$ stands for the Lebesgue measure, $M$ is the whole phase space, $A$
the attractor and $\varphi(x)$ and $\rho_0(x)$ are respectively a continuous
dynamical function and a normalized initial distribution function. 
Sinai, Ruelle and Bowen \cite{4} showed that the SRB measures for the axiom-A
systems satisfy Eq. (\ref{1.4}) as well. Hence, the
measure satisfying Eq. (\ref{1.4}) will be refered to as a physical measure.
Since the left hand side of Eq. (\ref{1.4}) is the average $\langle \varphi
\rangle_t$ of
$\varphi$ at time $t$, Eq. (\ref{1.4}) can be generalized to define a physical
measure
$\mu$ for systems with escape \cite{20,20b}

\begin{equation}
\lim_{t\to+\infty} \langle \varphi \rangle_t  = \lim_{t\to+\infty}
{\displaystyle \int_M
\varphi(S^tx) \rho_0(x) \mu_0(dx) \over \displaystyle \int_M
\chi_M(S^tx) \rho_0(x) \mu_0(dx)} = {\displaystyle
\int_{A'} \varphi(x)\mu(dx) \over \displaystyle  \int_{A'} \mu(dx)} \ , 
\label{1.5}
\end{equation}

\noindent
where $\chi_M$ stands for the characteristic function of the phase space
$M$ and $A'$ is the support of $\mu$. The denominator of the middle expression
is necessary as the total probability is not preserved. 
When there is no escape, the physical measure defined by Eq. (\ref{1.4}) is
supported by the attractor and, thus,  is invariant. On the other hand, when
there is escape, the physical measure defined by Eq. (\ref{1.5}) is, in general,
not an invariant measure.  However, since the ratio
$\int_{A'}\varphi(x)\mu(dx)/\int_{A'} \mu(dx)$ does not evolve in time, such a
measure is called conditionally invariant \cite{20,20b}.
We remark that, when they are of axiom-A type, the systems with escape also
possess ``natural'' invariant measures supported by the fractal repeller, which
are specified e.g., by a variational
principle\cite{1,14,19b,20,20b,21,22,43,HOY}. It is the natural invariant
measure that characterizes the ergodic properties such as the Lyapunov
exponents, but not the conditionally invariant physical measure defined by Eq.
(\ref{1.5}).

Now we revisit thermostated and open systems. 
As explained before, a thermostated system is dissipative because of
the fictitious damping force. Then, a nonequilibrium stationary state is
described by an asymptotic SRB measure defined by Eq. (\ref{1.2})
\cite{6,7,8,9,10,11,12,13,14}. 
For open chaotic Hamiltonian systems with a flux boundary condition, a
nonequilibrium stationary state obeying Fick's law is described
by a measure with a fractal structure along the contracting direction
\cite{15,16,17}. Since the measure is smooth along the expanding direction and
can be defined by Eq. (\ref{1.4}), it is an SRB measure\cite{15}.
In both cases, those SRB measures fully characterize the transport properties.
However, it should be noticed that the two cases are different because the
invariant set of an open system is a fractal repeller which is fractal in both
the stable and unstable directions and, thus, does not support an SRB measure,
while the invariant set of a thermostated system is an attractor which does
support an SRB measure.

By applying Eq. (\ref{1.4}) for an initial constant density, one obtains a
method of constructing an SRB measure for a map \cite{Ott}: 1) \ Approximate
the measure by iterating an initial constant density finite times, 2) \
calculate the average with its result and 3) \ take the limit of infinite
iterations.  Several methods are also proposed where unstable periodic orbits
or trajectory segments are used to write down an SRB measure and averages
with respect to it (cf. Refs. \cite{7,9,10,12,19,19b,EP} and references
therein).  However, it is not easy to evaluate the convergence rate of the
limits in Eqs. (\ref{1.4})-(\ref{1.5}), particularly for non-expanding maps, and
to 
extract
exponentially decaying terms from an averaged dynamical function $\langle
\varphi \rangle_t$, which are the Pollicott-Ruelle
resonances\cite{Po,Ru,PoRuRes,16,29,36,36b,37}. 
One of the reasons is that the long-time limit of the measure can be
defined only via the ensemble average as shown in  Eqs. (\ref{1.4}) and
(\ref{1.5}).  For non-expanding maps, the distribution function itself does not
have a well-defined long-time limit. In other words, an asymptotic SRB measure
is the {\it weak} limit of an initial measure.  
In an analytical construction of SRB measures which we shall explain for three
Baker-type maps, the weak convergence of measures is converted to the usual
convergence (technically speaking, the {\it strong} convergence) of partially
integrated distribution functions.  Contrary to the evolution equation of
a distribution function (the Frobenius-Perron equation), the evolution equation
of a partially integrated distribution is contractive and possesses a
well-defined long-time limit. This equation is similar to de Rham's functional
equation \cite{24}, which was introduced to deal with singular functions such as
continuous functions with zero derivatives almost everywhere, and its
contraction rate gives the convergence rate of the averaged dynamical function
$\langle
\varphi
\rangle_t$. In Sec. II, we review some properties of singular functions
including those of de Rham's functional equation. In Sec. III, an SRB measure is
constructed for a non-conservative reversible Baker map with the aid of de
Rham's equation. The model illustrates the two fundamental features of the
thermostated systems, namely the phase space contraction and time reversibility,
and we discuss the interrelation between the two features. In Sec. IV, a Baker
map with a Cantor-like invariant set is studied. When there is no escape, the
map possesses a strange attractor, which is a direct product of the unit
interval and a Cantor set. On the other hand, when there is escape, the map has
an invariant set, which is the direct product of two Cantor sets, and is a simple
example of an open system with an absorbing boundary condition. Physical
measures for the map with and without escape are constructed with the aid of de
Rham's equation and the natural invariant measure for the map with escape is
derived.  In Sec. V, we investigate the properties of a Baker map with
a flux boundary condition, which mimics a chemical-reaction dynamics with a flux
boundary condition (cf. Ref. \cite{ElKa}). We show that an SRB-type stationary
distribution describes the reaction dynamics and that the slowest relaxation to
it is characterized by a decay mode (i.e., the Pollicott-Ruelle resonance),
which is a conditionally invariant measure. Sec. VI is devoted to concluding
remarks. Technical details of the construction of measures are presented in
Appendixes. 

\section{SINGULAR FUNCTIONS AND DE RHAM EQUATION}

Basic tools of our analytical construction of SRB measures are singular functions
and de Rham's functional equation.
Singular functions such as the Weierstrass function or the Takagi function
were originally introduced as pathological counter examples to the intuitive
picture of functions. 
These singular functions play an important role in chaotic dynamics. A 
step towards the analytical treatment of
singular functions was given by de Rham \cite{24}, who showed many of them can
be characterized as a unique fixed point of a contraction mapping in a space
of functions. In this section, we briefly  review the properties of some
singular functions and the relation between them and chaotic dynamical systems
(cf. Refs.\cite{32} and \cite{38}).

The first example of a nowhere differentiable continuous function was given by
Weierstrass in 1872 \cite{25} (Fig.~\ref{fig1}):

\begin{equation}
W_{a,b}(x) = \sum_{n=0}^\infty a^n \ \cos(b^n \pi x) \ , \label{2.1}
\end{equation}

\noindent
where $0<a<1$, $b$ is a positive number and $ab \ge 1$. Moser \cite{26} used
the Weierstrass function to construct a nowhere differentiable torus for an
analytic Anosov system. Yamaguti and Hata \cite{27}
used it as a generating function
of orbits for the logistic map and discussed some generalizations. Also
Weierstrass functions are eigenfunctions of the Frobenius-Perron operator for
the Renyi map $Sx=rx$ (mod 1) with $r$ a positive integer \cite{29,28}.

In 1903, Takagi gave a simpler example of a nowhere differentiable continuous
function \cite{30} (Fig.~\ref{fig2}):

\begin{equation}
T(x) = \sum_{n=0}^\infty {1\over 2^n} \psi(2^n x) \ , \label{2.2}
\end{equation}

\noindent
where $\psi(x)=|x-[x+1/2]|$ and $[y]$ stands for the maximum integer which
does not exceed $y$. In 1930, van der Waerden gave a similar function which is
obtained from Eq. (\ref{2.2}) by replacing $2^n$ to $10^n$ \cite{31}. In 1957,
the Takagi function was rediscovered by de Rham \cite{24}. Some generalizations
of the Takagi function were discussed by Hata and Yamaguti \cite{27,32}. The
function 
$T(x)-1/2$ is the
eigenfunction of the Frobenius-Perron operator for the map $Sx=2x$ (mod 1) with
eigenvalue 1/2. Also, the Takagi function and related functions were found to
describe the nonequilibrium stationary state obeying Fick's law for the
multi-Baker map \cite{15}.

In the theory of the Lebesgue integral, there appear nonconstant 
functions with zero
derivatives almost everywhere, which are sometimes referred to as Lebesgue's
singular functions \cite{33,34}. One typical example is the Cantor function
(Fig.~\ref{fig3}). A more interesting example  $f_\alpha(x)$ ($0< \alpha<1,
\alpha\not=1/2$) is the unique function satisfying

\begin{equation}
f_\alpha(x)=\cases{\alpha f_\alpha(2x) \ , &$x\in [0,1/2]$ \cr
(1-\alpha) f_\alpha(2x-1)+\alpha \ , &$x\in [1/2,1]$ \cr} \ ,
\label{2.3}
\end{equation}

\noindent
which is strictly increasing and continuous, but has zero derivatives almost
everywhere with respect to the Lebesgue measure \cite{34} (Fig.~\ref{fig4}).
Note that such functions do not satisfy the fundamental theorem of
calculus \cite{33}:

$$
f_\alpha(1)-f_\alpha(0)=1\not=0=\int_0^1 f_\alpha'(x) dx \ . 
$$

\noindent
The function $f_\alpha(x)$ with real $\alpha$ represents a
cumulative distribution function of an ergodic measure for the
dyadic map $Sx=2x$ (mod 1) (cf. examples given on p.626 of Ref.\cite{2} and on
p.36 of Ref.\cite{35}). The eigenfunctionals of the Frobenius-Perron operator
for the multi-Bernoulli map and the multi-Baker map can be represented as the
Riemann-Stieltjes integrals with respect to $f_\alpha(x)$ with complex $\alpha$
\cite{36,36b}. In Ref.\cite{37}, it was shown that, for a class of
piecewise linear maps, the left eigenfunctionals of the
Frobenius-Perron operator admit a
representation in
terms of singular functions similar to $f_\alpha(x)$.
As pointed out by Hata and Yamaguti \cite{27,32}, $f_\alpha(x)$ is analytic with
respect to the parameter $\alpha$ though it is a singular function of $x$, and
there exists an interesting relation between
the parameter derivative of $f_\alpha(x)$ and the Takagi
function:

\begin{equation}
{d\over d\alpha} f_\alpha(x)\bigg|_{\alpha=1/2} = 2 T(x) \ . \label{Yamag}
\end{equation}

In 1957, de Rham found that the Weierstrass function, the Takagi
function and Lebesgue's singular function as well as other singular
functions are fully characterized as a unique solution
$f$ of a functional equation

\begin{equation}
f(x)={\cal F}f(x) + g(x) \ , \label{2.4}
\end{equation}

\noindent
where $g(x)$ is a given function and ${\cal F}$ is a contraction
mapping with ${\cal F} 0=0$ defined in the space of bounded
functions on the unit interval [0,1] \cite{24}.
Then, he generalized the functional equation (\ref{2.4}) to describe 
fractal continuous
curves such as the ones by Koch or L\'evy. The contraction mapping 
means that the
inequality

$$
{ \Norm{{\cal F}f_1 - {\cal F}f_2 } } \le \lambda 
{ \Norm{ f_1 - f_2 } } 
$$

\noindent
holds for some constant $0<\lambda<1$ and for any functions $f_1$ and
$f_2$, where the function norm ${ \Norm{\cdot} }$ is defined as the
supremum: ${ \Norm{f} }\equiv \sup_{[0,1]}|f(x)|$. 
The existence of a unique solution of de Rham's functional equation (\ref{2.4})
immediately follows from Banach's contraction mapping theorem \cite{34} 
and the fact that
the mapping $f\to {\cal F} f+g$ is contraction.
Let the mapping ${\cal F}_{\alpha,\beta}$ be

\begin{equation}
{\cal F}_{\alpha,\beta} f(x)\equiv \cases{\alpha f(2x) \ , &$x\in
[0,1/2]$ \cr \beta f(2x-1) \ , &$x\in (1/2,1]$
\cr} \ , \label{2.5}
\end{equation}

\noindent
then the Weierstrass function $W_{a,2}$, the Takagi function $T(x)$ 
and Lebesgue's
singular function $f_\alpha(x)$ are the unique solution of (\ref{2.4}) 
respectively for
$\alpha=\beta=a$, $g(x)=\cos \pi x$; for $\alpha=\beta=1/2$, $g(x)=|x-[x+1/2]|$;
and for $\beta=1-\alpha$, $g(x)=\alpha \theta(x-1/2)$ with $\theta$ the step
function \cite{24}. For more information on singular functions, see
Refs.\cite{32} and \cite{38}.

Before closing this section, we remark on Riemann-Stieltjes integrals with
respect to singular functions, which will appear in the next section. Suppose
$f(x)$ is of bounded variation and $\varphi(x)$ is continuous. Then, the
Riemann-Stieltjes integral of $\varphi$ with respect to $f$ is defined by the
limit

\begin{equation}
\int_a^b \varphi(x) df(x) \equiv \lim_{{\scriptscriptstyle \max(x_k-x_{k-1})}\to
0} \ \ \ \sum_{k=1}^n  \varphi(\xi_k) \{f(x_k)-f(x_{k-1})\} \ , \label{2.6}
\end{equation}

\noindent
where $\{x_k\}$ is a partition of [$a,b]$: $a=x_0<x_1<\cdots<x_n=b$, and
$x_{k-1}\le \xi_k \le x_k$ \cite{33,34}. Since the formula for the integration
by parts

$$
\int_a^b \varphi(x) df(x) + \int_a^b f(x) d\varphi(x) = \varphi(b)
f(b)-\varphi(a) f(a) \ , 
$$

\noindent
holds in general, the Riemann-Stieltjes integral of a function of bounded
variation with respect to a continuous function can be defined as above.
At first sight, the evaluation of the Riemann-Stieltjes integral (\ref{2.6}) 
seems to be
difficult. But, for a class of functions obeying de Rham's equation, it is not
the case. As an example, consider the Fourier-Laplace transformation of
$f_\alpha(x)$:

$$
I(\eta) \equiv \int_0^1 e^{i\eta x} df_\alpha(x) \ . 
$$

\noindent
By dividing the integral into the ones over [0,1/2] and [1/2,1] and changing the
variable $x$ to $x/2$ and $(x+1)/2$ respectively, we have the recursion relation

\begin{eqnarray}
I(\eta) &=& \int_0^{1} e^{i\eta x/2} df_\alpha(x/2)+  \int_0^{1}
e^{i\eta(x+1)/2} df_\alpha((x+1)/2) \nonumber \\
&=& \alpha\int_0^{1} e^{i\eta x/2} df_\alpha(x)+ (1-\alpha) e^{i\eta/2} 
\int_0^{1}
e^{i\eta x/2} df_\alpha(x) \nonumber \\
&=& \{ \alpha + (1-\alpha) e^{i\eta/2} \} I(\eta/2) \ , \nonumber
\end{eqnarray}

\noindent
where the de Rham equation (\ref{2.4}) is used in the second equality. Because
$I(0)=\int_0^1 df_\alpha(x)=f_\alpha(1)=1$, the above recursion relation gives
\cite{37,32,38,39}

\begin{equation}
\int_0^1 e^{i\eta x} df_\alpha(x) = I(\eta) = \prod_{n=1}^\infty \{ \alpha +
(1-\alpha) e^{i\eta/2^n} \}  \ . \label{2.7}
\end{equation}

\noindent
Note that, for $\alpha\not=1/2$, $I(2^m \pi)=I(2\pi)(\not=0)$ ($m=0,1,\cdots$)
and, hence, the Fourier-Laplace transformation of $f_\alpha$ does 
not satisfy the
Riemann-Lebesgue lemma: $\lim_{\eta\to\infty} I(\eta) \not=0$, that again
implies the singular nature of $f_\alpha$.

Further we notice that the formula (\ref{2.7}) and the Hata-Yamaguti relation
(\ref{Yamag}) relate the Lebesgue's singular function and the Takagi
function to the Weierstrass functions:

\begin{eqnarray}
f_\alpha(x) &=& \alpha - \sum_{s>0:{\rm odd}}{{\rm Im} I(2\pi s) \over \pi s}
\ W_{1/2,2}(2sx)+ \sum_{s>0:{\rm odd}}{{\rm Re} I(2\pi s) -1 \over \pi s}
\ {\widetilde W}_{1/2,2}(2sx) \ , \nonumber \\
T(x) &=& {1\over 2} - \sum_{s>0:{\rm odd}}{2 \over \pi^2 s^2 }
\ W_{1/2,2}(2sx) \ , \nonumber
\end{eqnarray}

\noindent
where the sums runs over positive odd integers, $W_{1/2,2}(x)$ is the
Weierstrass function (\ref{2.1}) and ${\widetilde W}_{1/2,2}(x)$ is a singular
function obtained from (\ref{2.1}) by replacing $\cos$ to $\sin$.

\vskip 1truecm

\section{SRB MEASURE FOR A NON-CONSERVATIVE REVERSIBLE BAKER MAP}

In thermostated systems, dynamics is non-conservative due to the damping force
mimicking a heat reservoir and thus, the forward time evolution is different
from the backward time evolution. However, it is time reversible
\cite{6,7,8,9,10,11,12,13,14}.
It is therefore interesting to see how these two features are compatible, that
we shall study with a simple map.
One of the simplest non-conservative systems which have time 
reversal symmtery is given by (Fig.~\ref{fig5})

\begin{equation}
\Phi(x,y) = \cases{ \bigl(x/l,r \ y \bigr) \ , &$x\in [0,l]$ \cr
\bigl((x-l)/r,l \ y+r\bigr) \ , &$x\in (l,1]$ \cr} \label{3.1}
\end{equation}

\noindent
where $l+r=1$ and $0<l<1$. The map is non-conservative since its Jacobian
takes $r/l$ for $x\in [0,l]$ and $l/r$ for $x\in (l,1]$,
both of which are different from 1. But the map has a time reversal symmetry
represented by an involution $I(x,y)\equiv (1-y,1-x)$: $I\Phi I=\Phi^{-1}$.
The Frobenius-Perron equation governing the time evolution of distribution
functions (with respect to the Lebesgue measure) is given by

\begin{eqnarray}
\rho_{t+1}(x,y)&=&U\rho_{t}(x,y)\equiv \int dx' dy'
\delta\bigl[(x,y)-\Phi(x',y')\bigr] \rho_{t}(x',y') \nonumber \\
&=& \cases{ {\displaystyle l \over \displaystyle r} \ \rho_t\bigl(l \ x,
y/r \bigr) \ , &$y\in [0,r]$ \cr
{\displaystyle r \over \displaystyle l} \ \rho_t\bigl(r \
x+l,(y-r)/l\bigr) \ , &$y\in (r,1]$ \cr} \label{3.2}  
\end{eqnarray}

\noindent
where $U$ stands for the Frobenius-Perron operator defined by the second
equality and $\delta[\cdot]$ is the two-dimensional delta function. Since the map
$\Phi$ is not conservative, the numerical factors $r/l$ and $l/r$
different from 1 appear in the last expression.

\subsection{An SRB measure for the forward time evolution}

First we explain our algorithm to construct SRB measures and apply it to the
forward time evolution. Our goal is to show that an expectation value of
the dynamical function $\varphi(x,y)$ with respect to $\rho_t$ converges
for $t\to +\infty$ to an expectation with respect to a singular measure given
below, when the initial distribution function $\rho_0$ is continuously
differentiable in $x$, and a dynamical function $\varphi$ is continuously
differentiable in $y$ and continuous in $x$. We remark that the convergence 
rate is controlled by the smoothness of the initial distribution function and 
the dynamical function, and the condition given above is sufficient for the
exponential convergence.  

The first step in our explicit construction of the singular measure is to
express the expectation value by the partially integrated distribution
function, i.e.,

\begin{eqnarray}
\int_{[0,1]^2} &\varphi&(x,y) \rho_{t}(x,y) dx dy = \int_{[0,1]^2} \varphi(x,y)
\ d_yP_{t}(x,y) dx\nonumber \\
&=& \int_0^1 \varphi(x,1) \
P_{t}(x,1)  dx - \int_{[0,1]^2} \partial_y\varphi(x,y) \
P_{t}(x,y) dx dy
 \ , \label{3.3a} 
\end{eqnarray}

\noindent
where $P_{t}(x,y)\equiv \int_0^y dy' \rho_t(x,y')$ is the partially integrated
distribution function, the symbol $d_y$ stands for the 
Riemann-Stieltjes integral of
$P_t$ only with respect to the variable $y$ \cite{42} and the integration 
by parts is used in
the second equality. The evolution equation for $P_t$ can be obtained 
easily from
Eq. (\ref{3.2}):

\begin{equation}
P_{t+1}(x,y)= \cases{ l \ P_t\bigl(l \ x, y/r \bigr) \ , &$y\in [0,r]$
\cr  r \ P_t\bigl(r \ x+l,(y-r)/l\bigr) + l \ P_t\bigl(l \
x,1\bigr)  \ . &$y\in (r,1]$ \cr} \label{3.3b}
\end{equation}

\noindent
Partial integration of the distribution function changes the prefactors from
$r/l$ and $l/r$ respectively to $r$ and $l$, which are strictly less
than unity. Because of this, the evolution equation (\ref{3.3b}) is similar to
de Rham's functional equation (\ref{2.4}).

The next step in our construction of the singular measure is to calculate
the long time limit of $P_t$. Putting
$y=1$ in Eq. (\ref{3.3b}), we obtain

\begin{equation}
P_{t+1}(x,1)= r \ P_t\bigl(r \ x+l,1\bigr) + l \ 
P_t\bigl(l \ x,1\bigr)
\ . \label{3.4}
\end{equation}

\noindent
Note that this is nothing but the Frobenius-Perron equation for a one-dimensional
chaotic map (strictly speaking, an exact map cf. \cite{LaMa})
$Sx=x/l$ (for
$x\in [0,l]$) and $Sx=(x-l)/r$ (for $x\in (l,1]$), which admits the
Lebesgue measure as the invariant measure. Therefore, the normalization
integral $\int^1_0 dx P_t(x,1)$ is invariant:

$$
\int^1_0 dx P_t(x,1) = \int^1_0 dx P_{t-1}(x,1) = \cdots = \int^1_0 dx P_0(x,1)
\ ,
$$

\noindent
and is equal to the long time limit of the partially integrated distribution
function $P_t(x,1)$ \cite{LaMa}.  As will be shown in Appendix A, the
convergence rate is $\lambda$.

\begin{equation}
P_t(x,1)= \int_0^1 dx' P_0(x',1) + {\rm O}(\lambda^t)
=\int_{[0,1]^2} \rho_0(x',y') dx'dy' + {\rm O}(\lambda^t) \ . \label{3.5}
\end{equation}

In order to proceed with the calculation, one needs the following lemma.
(For its proof, see Appendix A.)

\vskip 10 pt

\noindent{\bf Lemma} \ \ \ {\it
Let ${\cal F}$ be a linear contraction mapping with the contraction constant
$0<\lambda<1$. And let $g_t\equiv g^{(0)}+\nu^t g^{(1)} 
+g^{(2)}_t$ be a
given function where 
$\nu$ is a constant satisfying $\lambda<|\nu|\le 1$, 
and $g^{(2)}_t= {\rm O}(\lambda^t)$. Then, the solution of
the functional equation
\begin{equation}
f_{t+1} = {\cal F} f_t + g_t \ , \label{3.6}
\end{equation}
is given by
\begin{equation}
f_t = f_\infty^{(0)} + \nu^t f_\infty^{(1)} + {\rm O}(t \lambda^t) \ ,
\label{3.7a}
\end{equation}
where $f_\infty^{(0)}$ and $f_\infty^{(1)}$ are the unique solutions of the
following fixed point equations
\begin{eqnarray}
f_\infty^{(0)} &=& {\cal F} f_\infty^{(0)} + g^{(0)} \ , \label{3.7b} \\
f_\infty^{(1)} &=& {1\over \nu}{\cal F} f_\infty^{(1)} + {g^{(1)}\over \nu} \ .
\label{3.7c}
\end{eqnarray}
}

Now we go back to the equation (\ref{3.3b}) for $P_t(x,y)$, which can be
rewritten as

\begin{equation}
P_{t+1}={\cal F} P_t + g_t \ , \label{3.8a}
\end{equation}

\noindent
where the contraction mapping $\cal F$ and a function $g_t$ are respectively 
given by

\begin{equation}
{\cal F} P(x,y) = \cases{ l \ P\bigl(l \ x, y/r \bigr) \ , &$y\in [0,r]$
\cr  r \ P\bigl(r \ x+l,(y-r)/l\bigr)  \ . &$y\in (r,1]$ \cr}
\label{3.8b} 
\end{equation}

\noindent
and

\begin{equation}
g_t(x,y)\equiv \cases{ 0 \ , &$y\in [0,r]$
\cr  l \ P_t\bigl(l \
x,1\bigr)  \ , &$y\in (r,1]$ \cr} =
{\bar g}^{(0)}(y) \ \displaystyle \int_{[0,1]^2} \rho_0(x',y') dx'dy' +
{\rm O}(\lambda^t) \ , \label{3.8c}
\end{equation}

\noindent
with

\begin{equation}
{\bar g}^{(0)}(y)=\cases{ 0 \ , &$y\in [0,r]$ \cr l  \ . 
&$y\in (r,1]$ \cr}
\label{3.8d}
\end{equation}

\noindent
The contraction constant of the mapping $\cal F$ is $\lambda
={\rm max}(l,r)$ 
and Eq. (\ref{3.5}) is used to derive the left-hand side of
Eq. (\ref{3.8c}). 
As the equations (\ref{3.8a}), (\ref{3.8b}), (\ref{3.8c})
and (\ref{3.8d}) satisfy the condition of the lemma and the contraction mapping 
$\cal F$ given by (\ref{3.8b}) is linear, the lemma implies

\begin{equation}
P_t(x,y)= F_{l}(y)\ \displaystyle \int_{[0,1]^2}
\rho_0(x',y') dx'dy' + {\rm O}(t \lambda^t) \ , \label{3.9a}
\end{equation}

\noindent
where $F_{l}(y)$ is the unique solution of de Rham's functional equation

\begin{equation}
F_{l}(y) = {\cal F} F_{l}(y) +{\bar g}^{(0)}(y)  = \cases{ l \ 
F_{l}\bigl(y/r \bigr) \ , &$y\in [0,r]$ \cr 
r \ F_{l}\bigl((y-r)/l\bigr) + l \ . &$y\in (r,1]$ \cr} \label{3.9b}
\end{equation}

By substituting Eq. (\ref{3.9a}) into Eq. (\ref{3.3a}) and employing the
integration by parts, we have

\begin{equation}
\int_{[0,1]^2} \varphi(x,y) \rho_{t}(x,y) dx dy = \int_{[0,1]^2} \varphi(x,y) \
dx dF_{l}(y)\ \int_{[0,1]^2} \rho_0(x',y') dx'dy' + {\rm O}(t \lambda^t) \ .
\label{3.10}
\end{equation}

\noindent
We remind the reader that Eq. (\ref{3.10}) holds for any continuous function 
$\varphi(x,y)$ and
any integrable function $\rho_0(x,y)$ which are continuously differentiable
respectively in $y$ and $x$. If $\rho_0$ is normalized with respect to
the Lebesgue measure, Eq. (\ref{3.10}) gives

\begin{equation}
\lim_{t\to \infty} \int_{[0,1]^2} \varphi(x,y) \rho_{t}(x,y) dx dy =
\int_{[0,1]^2} \varphi(x,y) \ dx dF_{l}(y) \ . \label{3.11}
\end{equation}

\noindent
This shows that the physical measure $\mu_{\rm ph}$ of the system is given by

\begin{equation}
\mu_{\rm ph}\biggl([0,x)\times [0,y)\biggr)= x F_{l}(y) \ . \label{3.12}
\end{equation}

\noindent
Clearly it is absolutely continuous
with respect to the Lebesgue measure
along the expanding $x$-direction and, thus,
is an SRB measure. As studied in Ref.\cite{37}, the function $F_{l}$ is
non-decreasing, has zero derivatives almost everywhere
except for $r=l=1/2$
with respect to the
Lebesgue measure and is H\"older continuous with exponent $\delta=-\ln
\max(l,r)/\ln \min(l^{-1},r^{-1})=1$ (i.e., $|F_{l}(x)-F_{l}(y)|\le
A|x-y|$). The graph of $F_{l}$ is a fractal (Fig. 6), but its 
fractal dimension is
$D=1$ as a result of the Besicovich-Ursell inequality \cite{40}: 
$1\le D \le 2-\delta=1$.
Moreover, the one-dimensional measure defined by $F_{l}$ is a multifractal 
two-scale Cantor measure, the dimension spectrum $D_q$ ($-\infty <q< +\infty$) 
of which is given as the  solution of \cite{37,2CAN}

$$
{l^q \ r^{(1-q) D_q} }+{r^q \ l^{(1-q) D_q} }=1 \ . 
$$

\noindent
There exists an interesting relation between $F_{l}$ and Lebesgue's singular
function $f_\alpha$:

$$
F_{l}(y)=f_{l}\circ f_{r}^{-1}(y) \ , 
$$ 

\noindent
which immediately follows from the fact that the right-hand side obeys the same
functional equation as $F_{l}$.  Note that, since $f_{r}$ is continuous and
strictly increasing, the inverse $f_{r}^{-1}$ exisits and is also strictly
increasing. As a composite function of two strictly increasing functions $f_{l}$
and $f_{r}^{-1}$, $F_{l}$ is also strictly increasing. 
Because of these singular properties of $F_{l}$, the physical measure $\mu_{\rm
ph}$ is singular along the contracting $y$-direction.

The physical measure $\mu_{\rm ph}$ is mixing with respect to the map $\Phi$.
Indeed, by considering $P_t(x,y)=\int_0^y \rho_t(x,y') dF_{l}(y')$ instead of
$P_t(x,y)=\int_0^y \rho_t(x,y') dy'$ and following exactly the same arguments
as above, one obtains

\begin{equation}
\lim_{t\to \infty} \int_{[0,1]^2} \varphi(x,y) \rho_{t}(x,y) dx dF_{l}(y) =
\int_{[0,1]^2} \varphi(x,y) \ dx dF_{l}(y) \ , \label{3.13}
\end{equation}

\noindent
provided that $\varphi(x,y)$ and $\rho_0(x,y)$ are continuously differentiable
respectively in $y$ and $x$ and that $\rho_0$ is normalized: $\int
\rho_0(x,y) dx dy =1$. 
By using this fact, the Lyapunov exponents can be
analytically calculated as follows: The Jacobian matrix for
$\Phi$ is diagonal, and the logarithm of the component along the expanding $x$
direction,  the local expanding rate $\Lambda_x (x, y)$, is

$$
\Lambda_x(x,y)=\cases{-\ln l \ , &$x\in [0,l]$ \cr
-\ln r \ . &$x\in (l,1]$ \cr} 
$$

\noindent
The Lyapunov exponent along the $x$ direction (the positive Lyapunov 
exponent) is defined as the average of $\Lambda_x (x, y)$ over an orbit starting
from some initial point $(x,y).$ 
Since the system is ergodic, the Lyapunov exponent can be obtained from 
Eq. (\ref{1.2}) using the measure $\mu_{\rm ph}$ :

\begin{equation}
\lambda_x(\Phi, \mu_{\rm ph}) \equiv \lim_{T\to +\infty} {1\over T}\sum_{t=0}^T
\Lambda_x\Bigl(\Phi^t(x,y) \Bigl) = \int_{[0,1]^2}\Lambda_x(x,y) dx dF_{l}(y)
= -l \ln l - r \ln r \ . \label{3.14a}
\end{equation}

\noindent
Similarly, the Lyapunov exponent along the $y$ direction (the negative Lyapunov
exponent) is  

\begin{equation}
\lambda_y(\Phi, \mu_{\rm ph}) = l \ln r + r \ln l \ . \label{3.14b}
\end{equation}

The sum of the two Lyapunov exponents is negative :

$$
\lambda_x(\Phi, \mu_{\rm ph}) + \lambda_y(\Phi, \mu_{\rm ph}) =
(r-l) \ \ln\left({l\over r}\right) <0 \ .
$$

\noindent
Hence, areas are contracted on average by the map $\Phi$ and the map possesses an
attractor $A$. From  Eq. (\ref{3.11}), one finds that the attractor $A$ is the
support of the SRB measure $\mu_{\rm ph}$. When $l\not= 1/2$, the two-dimensional Lebesgue
measure of the attractor $A$ is zero since the function $F_l$ has zero
derivatives almost everywhere with respect to the Lebesgue measure. Moreover,
according to  Young's formula \cite{Young} (which is the Kaplan-Yorke formula
\cite{2,3,Ott,41} for two-dimensional ergodic systems), the information
dimension of $A$ is given by :

$$ 
D_{\rm I} = 1 + {\lambda_x(\Phi, \mu_{\rm ph})\over|\lambda_y(\Phi, \mu_{\rm
ph})|} = 1 +
\left|{l \ln l + r \ln r \over  l \ln r + r \ln l}\right| < 2 \ , 
$$

\noindent
which is less than two. Therefore, $A$ is a fractal set. 
On the other hand, since the function $F_l$ is strictly increasing, for any
non-empty rectangle $[x_0,x_0+\epsilon)\times [y_0,y_0+\epsilon')$ ($\epsilon
>0$ and $\epsilon'>0$), we have 

$$
\mu_{\rm ph}\biggl([x_0,x_0+\epsilon)\times [y_0,y_0+\epsilon')\biggr)= 
\epsilon \biggl\{ F_{l}(y_0+\epsilon') - F_{l}(y_0) \biggr\} > 0 \ ,
$$

\noindent
which implies that $A \cap [x_0,x_0+\epsilon)\times [y_0,y_0+\epsilon')\not=
\emptyset$ and, hence, the attractor $A$ is dense in the whole phase space
$[0,1)^2$. This phase-space structure is in contrast to the one of a dissipative
system usually studied \cite{Ott} (see also the next section), where an
attractor is a Cantor-like set. 

\subsection{An SRB measure for the backward time evolution}

Now we consider the backward time evolution. In the same way as the forward
evolution, we find that another partially integrated distribution function
${\bar P}_t(x,y)=\int_0^x dx' \rho_t(x',y)$ converges for $t\to -\infty$ and
we have

\begin{equation}
\int_{[0,1]^2} \varphi(x,y) \rho_{t}(x,y) dx dy = \int_{[0,1]^2} \varphi(x,y) \
d{\bar F}_{r}(x)dy \ \int_{[0,1]^2} \rho_0(x',y') dx'dy' + {\rm O}(|t|
\lambda^{|t|}) \ , \label{3.15} 
\end{equation}

\noindent
where $t=0,-1,-2,\cdots$ and a singular function ${\bar F}_{r}$ is given by
${\bar F}_{r}(x)=1-F_{l}(1-x)$. As before, Eq. (\ref{3.15}) implies that the
asymptotic physical measure ${\bar \mu}_{\rm ph}$ is given by
${\bar \mu}_{\rm ph}([0,x)\times [0,y)) \equiv {\bar F}_{r}(x) y$. 
The measure ${\bar \mu}_{\rm ph}$ is then absolutely continuous with respect to
the Lebesgue measure along $y$ direction and singular along $x$ direction. This
correponds to the fact that the expanding and contracting directions are
interchanged for the backward motion. Actually, the measure ${\bar \mu}_{\rm ph}$
is precisely the one obtained from $\mu_{\rm ph}$ via the time 
reversal operation
$I$: ${\bar \mu}_{\rm ph}= I\mu_{\rm ph}$.  The measure ${\bar \mu}_{\rm ph}$ is
again mixing with respect to the backward time evolution $\Phi^{-t}$
($t=0,1,\cdots$). And the Lyapunov exponents are calculated as the phase space
averages of the local scaling rates for the inverse map $\Phi^{-1}$ with respect
to ${\bar \mu}_{\rm ph}$. 
For example, the positive Lyapunov exponent is the ${\bar \mu}_{\rm
ph}$-average of the local expanding rate: ${\bar
\Lambda}_y(x,y)= -\theta(r-y) \ln r -\theta(y- r) \ln l$ and
is equal to that for the original map $\Phi$. 

\begin{equation}
\lambda_y(\Phi^{-1},{\bar \mu}_{\rm ph}) =\int_{[0,1]^2}{\bar \Lambda}_y(x,y)
d{\bar F}_{r}(x)dy = -r \ln r -l \ln l
=\lambda_x(\Phi, \mu_{\rm ph}) \ . \label{3.16a}
\end{equation}

\noindent
The negative Lyapunov exponents of the two maps are also the same: 

\begin{equation}
\lambda_x(\Phi^{-1},{\bar \mu}_{\rm ph}) =
\lambda_y(\Phi, \mu_{\rm ph}) \ . \label{3.16b}
\end{equation}

\noindent
The equality
of Lyapunov exponents for ($\Phi,{\mu}_{\rm ph})$ and 
$(\Phi^{-1},{\bar \mu}_{\rm
ph})$ is a general consequence of the time reversal symmetry of the system.

We notice that the natural invariant measure ${\bar \mu}_{\rm
ph}$ of $\Phi^{-1}$ is also invariant under $\Phi$ as follows 
straightforwardly from the reversibililty of $\Phi$ :

$$
{\bar \mu}_{\rm ph}\bigl(\Phi^{-1}\{[0,x)\times [0,y)\}\bigr) =
{\bar \mu}_{\rm ph}\bigl([0,x)\times [0,y)\bigr) \ , 
$$

\noindent
is equivalent to

\begin{equation}{\bar \mu}_{\rm ph}\bigl([0,x)\times [0,y)\bigr) = {\bar
\mu}_{\rm ph}(\Phi\{[0,x)\times [0,y)\}) \ . \label{3.17}
\end{equation}

\noindent
That is, we may think of ${\bar \mu}_{\rm ph}$ as a repelling measure for $\Phi$,
in the sense that, while it is indeed invariant, any slight deviations from this
measure, if they are absolutely continuous with respect to the Lebesgue measure,
will evolve toward the measure $\mu_{\rm ph}$ for the attractor for $\Phi$ (cf.
Eq. (\ref{3.11})).
In short, we find, for a non-conservative reversible map $\Phi$, different SRB
measures ${\mu}_{\rm ph}$ and ${\bar \mu}_{\rm ph}$ for the forward and backward
time evolutions, respectively. And for each time evolution, one plays a role of
an attracting measure and the other a role of a repelling measure in the sense
just explained. This observation is a key element of the compatibility between
dynamical reversibility and irreversible behavior of statistical ensembles.
Indeed, when the dynamics is reversible and statistical ensembles
irreversibly approach a stationary ensemble ${\mu}_{\rm ph}$ for $t\to +\infty$,
there should exist another stationary ensemble ${\bar \mu}_{\rm ph}$ which is
obtained from ${\mu}_{\rm ph}$ by the time reversal operation. 
However, the new stationary ensemble ${\bar \mu}_{\rm ph}$ is repelling
and, thus, its existence is compatible with the irreversible behavior of
statistical ensembles. 

Further, the measure ${\bar \mu}_{\rm ph}$ is mixing with respect to the
forward time evolution $\Phi^t$ ($t=0,1,\cdots$).  Then it is interesting to
investigate the relation between Lyapunov exponents and the Kolmogorov-Sinai
(KS) entropy for $(\Phi,{\bar \mu}_{\rm ph})$. The Lyapunov exponents for
$(\Phi,{\bar \mu}_{\rm ph})$ are easily found to be

\begin{equation}
\lambda_x(\Phi,{\bar \mu}_{\rm ph}) =  -r \ln l -l \ln r
\ , \label{3.18a}
\end{equation}

\noindent
and

\begin{equation}
\lambda_y(\Phi,{\bar \mu}_{\rm ph}) =  r \ln r + l \ln l
\ . \label{3.18b}
\end{equation}

\noindent
It is useful to note that the positive (negative) Lyapunov exponent of
${\bar \mu}_{\rm ph}$ can be found by changing the sign of the
negative (positive) Lyapunov exponent of ${\mu}_{\rm ph}$. This fact
is widely used in the analysis of the Lyapunov spectrum of
thermostated many particle systems \cite{6,9,10}. In our case this result
follows from the observations that

\begin{equation}
\lambda_x(\Phi,{\bar \mu}_{\rm ph}) = - \lambda_y(\Phi,\mu_{\rm
ph}) 
\ , \label{3.19a}
\end{equation}

\noindent
and

\begin{equation}
\lambda_y(\Phi,{\bar \mu}_{\rm ph}) = - \lambda_x(\Phi,\mu_{\rm
ph}) \ . \label{3.19b}
\end{equation}

\noindent
Thus the Lyapunov spectrum changes sign under the exchange of 
$\mu_{\rm ph}$ and ${\bar \mu}_{\rm ph}.$
Now we turn to a calculation of the KS-entropy. First we note that the
KS-entropy of $(\Phi,{\mu}_{\rm ph})$ is

\begin{equation}
h_{KS} (\Phi,{\mu}_{\rm ph}) = \lambda_x(\Phi,\mu_{\rm ph}) \ ,
\label{3.20}
\end{equation}

\noindent
which follows from the Pesin's identity \cite{2}, since ${\mu}_{\rm
ph}$ is the SRB measure for the map $\Phi$, and which can also be computed
directly  from the entropy of the generating partition formed by the two
elements $0 \le x \le l$ and $l \le x \le 1.$ 
From the same partition, it is readily seen that the entropy of $(\Phi,{\bar
\mu}_{\rm ph})$ is also 

\begin{equation}
h_{KS} (\Phi,{\bar \mu}_{\rm ph}) = \lambda_x(\Phi,\mu_{\rm ph})
\ . \label{3.21}
\end{equation}

\noindent
Then the difference between the positive Lyapunov exponent and the KS-entropy
for $(\Phi,{\bar \mu}_{\rm ph})$ is

\begin{equation}
\lambda_x(\Phi,{\bar \mu}_{\rm ph}) - h_{KS} (\Phi,{\bar \mu}_{\rm ph})
= - \lambda_y(\Phi,\mu_{\rm ph}) - \lambda_x(\Phi,\mu_{\rm ph}) \ge 0
\ , \label{RueIneq}
\end{equation}

\noindent
which is strictly positive for $l\not=1/2$ since the right-hand side is
just the phase space contraction rate. Therefore, the mixing system $(\Phi,{\bar
\mu}_{\rm ph})$ violates Pesin's identity, but satisfies Ruelle's inequality
\cite{2}, as it should be since the measure ${\bar \mu}_{\rm ph}$ is not
an SRB measure for $\Phi$. 


\vskip 1truecm

\section{SRB MEASURE FOR A BAKER MAP WITH A CANTOR-LIKE INVARIANT SET}


Transport properties are also studied for open Hamiltonian systems with a flux
boundary condition \cite{15,16,17} or with an absorbing boundary condition
\cite{14,18,19,19b,20,20b,21,22}. Breymann,  T\'el and Vollmer used an open
non-conservative system to study an interrelation between
the thermostated systems approach and the open systems approach \cite{23}.
Further, a model used by Kaufmann, Lustfeld, N\'emeth and Sz\'epfalusky
\cite{KLNS} to investigate deterministic transient diffusion is 
a non-conservative open system. 
One of the important features of those open systems is the existence of
escape. So we investigate how escape affects the physical measure, by using a
Baker map with a Cantor-like invariant set \cite{12,Ott,41}.
The map is defined on the unit square $[0,1]^2$ (Fig.~\ref{fig7}):

\begin{equation}
\Psi(x,y) = \cases{ \bigl(x/l, \Lambda_1 y \bigr) \ , &$x\in [0,l]$ \cr
\bigl( (x-a)/r, \Lambda_2 y+ b \bigr) \ , &$x\in [a,a+r]$ \cr}
\label{4.1}
\end{equation}

\noindent
where $0< l \le a$, $0< r \le 1-a$, $0< \Lambda_1 \le b$ and 
$0< \Lambda_2 \le 1-b$. Here we introduce an escape for points $x \in (l, a)
\cup (a+r,1]$ and an inhomogeneity. 
For $\Lambda_1<l$ and $\Lambda_2>r$, or for $\Lambda_1>l$ and
$\Lambda_2<r$, the map $\Psi$ is partially attractive and partially repelling
 as discussed in the previous section and, for $\Lambda_1=r$, $\Lambda_2=l$, and
$l+r=1$, it is reduced to the previous model. 
It is everywhere attractive for $\Lambda_1<l$ and $\Lambda_2<r$, is conservative
for $\Lambda_1=l$ and $\Lambda_2=r$ and is everywhere repelling for
$\Lambda_1>l$ and $\Lambda_2>r$. 
Note that the last case is possible only when there exists an escape: $l+r<1$.

We show that when the initial distribution function $\rho_0$ is 
continuously
differentiable in $x$, and a dynamical function $\varphi$ is continuously
differentiable in $y$ and continuous in $x$, an expectation value of the
dynamical function $\varphi(x,y)$ with respect to $\rho_t$ decays exponentially:

\begin{equation}
\int_{[0,1]^2} \varphi(x,y) \rho_{t}(x,y) dx dy = {\nu}^t \int_{[0,1]^2}
\varphi(x,y) \ dx dG(y) \int_{[0,1]^2}
\rho_0(x,y) \ dH(x) dy + {\rm O}(t{\lambda'}^t)  
\ , \label{4.2}
\end{equation}

\noindent
where the decay rate is equal to the remainder volume per iteration:
$\nu \equiv  l+r$, $\lambda'=\max(l,r) (\le \nu$).
The functions $G$ and $H$ 
are (possibly) singular
functions defined as the unique solutions of de Rham equations :

\begin{equation}
G(y) = \cases{ {\displaystyle l \over \displaystyle l+r} 
G\bigl({\displaystyle y \over \displaystyle \Lambda_1} \bigr) \ , &$y\in
[0,\Lambda_1]$ \cr  
{\displaystyle l \over \displaystyle l+r} \ ,
&$y\in (\Lambda_1, b)$ \cr 
{\displaystyle r\over \displaystyle l+r}G\bigl({\displaystyle y-b
\over  \displaystyle \Lambda_2}\bigr) + {\displaystyle l \over \displaystyle
l+r} \ , &$y\in [b,\Lambda_2+b]$ \cr
1 \ . &$y\in (\Lambda_2+b,1]$ \cr 
} \label{4.10b} 
\end{equation}

\noindent
and

\begin{equation}
H(x) = \cases{ {\displaystyle l\over \displaystyle l+r}
H\biggl({\displaystyle x \over \displaystyle l} \biggr) \ ,
&$x\in [0,l]$ \cr 
{\displaystyle l\over \displaystyle l+r} \ ,
&$x\in (l, a)$ \cr 
{\displaystyle r\over \displaystyle l+r}
H\biggl({\displaystyle x-a \over \displaystyle r} \biggr) + 
{\displaystyle l \over \displaystyle l+r} \
, &$x\in [a,a+r]$ \cr 
1 \ . &$x\in (a+r,1]$ \cr 
} \label{4.7}
\end{equation}

\noindent
The derivation of Eq. (\ref{4.2}) is outlined in Appendix B.
Now we investigate the implications of Eq. (\ref{4.2}) in cases without and with
escape separately.


\vskip .7 truecm

\subsection{Baker map without escape}

Consider first the case where there is no escape ($l+r=1$). The unit square 
is then contracted onto a set which is a direct
product of a Cantor set in the $y$ direction and the unit interval,  $[0,1]$,
in the $x$  direction.  This direct product set is nothing but the strange
attractor of $\Psi$, 
whose Hausdorff dimension $1< H_D <2$.
In this case, $H(x)=x$ and (\ref{4.2}) reduces to an
expression similar to (\ref{3.15}) for the previous example.
And the function $G$ defines an invariant mixing measure $\mu_{\rm ph}$;

\begin{equation}
\mu_{\rm ph}\Bigl([0,x)\times [0,y)\Bigr)\equiv x G(y) \ . \label{4.11}
\end{equation}

\noindent
The measure $\mu_{\rm ph}$ is the one which provides 
long-term averages of dynamical functions and thus, is a physical measure.
Indeed, when $\int \rho_0(x,y) dxdy=1$, (\ref{4.2}) gives 

\begin{equation}
\bra \varphi \ket_t \equiv \int_{[0,1]^2} 
\varphi(x,y) \rho_{t}(x,y) dx dy  = \int_{[0,1]^2}
\varphi(x,y) \ dx dG(y) + {\rm
O}\left(t\left\{{\lambda'}\right\}^t\right)   \ . \label{4.12}
\end{equation}

\noindent
Since $\mu_{\rm ph}$ is smooth (more precisely absolutely 
continuous 
with respect to the Lebesgue measure) along
the expanding coordinate $x$, it is an SRB measure. As shown 
in Fig.~\ref{fig8}, the graph of $G(y)$ is a typical devil's staircase and the
measure $\mu_{\rm ph}$ is singular.
Note that the support of $\mu_{\rm ph}$ is the strange
attractor of $\Psi$.

\vskip .7 truecm

\subsection{Baker map with escape}

When $l+r<1$, almost all the points escape the 
unit square and there appears a fractal repeller,  which is
singular both in expanding and contracting directions.
Then, the invariant measure supported by the fractal repeller
is different from the physical measure defined by Eq. (\ref{1.5}), which is not
invariant 
but conditionally invariant under $\Psi$ \cite{20,20b}.

First we consider the physical measure. The formula (\ref{4.2}) holds even when
$l+r<1$. By setting $\varphi \equiv 1$ in (\ref{4.2}), the renormalization
factor
$\int \rho_t(x,y) dxdy$ is found to be

$$
\int_{[0,1]^2} \rho_{t}(x,y) dx dy = {\nu}^t \int_{[0,1]^2}
\rho_0(x,y) \ dH(x) dy + {\rm O}(t{\lambda'}^t)  \ .
$$

\noindent
Hence, the expectation value of $\varphi$ at time $t$ is

\begin{equation}
\bra \varphi \ket_t \equiv{\displaystyle \int_{[0,1]^2} 
\varphi(x,y) \rho_{t}(x,y)
dx dy \over \displaystyle \int_{[0,1]^2} \rho_{t}(x,y) dx dy} = \int_{[0,1]^2}
\varphi(x,y) \ dx dG(y) + {\rm
O}\left(t\left\{{\lambda'\over\nu}\right\}^t\right)   \ . \label{4.13}
\end{equation}

\noindent
This implies that the physical measure defined by Eq. (\ref{1.5}) is identical
to that for the Baker map without escape:

\begin{equation}
\mu_{\rm ph}\Bigl([0,x)\times [0,y)\Bigr) = x G(y) \ , \label{4.14}
\end{equation}

\noindent
provided that the ratio $l/r$ is common in two cases. As in the previous
case, the physical measure $\mu_{\rm ph}$ is singular and is supported by a
direct product of a Cantor set $C_1$ along $y$ and the unit interval 
along $x$: $[0,1]\times C_1$, which is the unstable manifold of the fractal
repeller. Since it is smooth along the expanding
$x$-direction, it is an SRB-like measure (in the sense that, although not
invariant, it is smooth along the expanding direction).

We observe that the support of $\mu_{\rm ph}$ 
is the unstable manifold of the repeller 
and is not an invariant set with respect to $\Psi$. Accordingly, the
physical measure 
$\mu_{\rm ph}$ is not an invariant measure. Indeed, the measure
$\mu_{\rm ph}$ satisfies

\begin{equation}
\mu_{\rm ph}\biggl(\Psi^{-1}\bigl\{[0,x)\times[0,y)\bigr\}\biggr)
= (l+r) \ \mu_{\rm ph}\biggl([0,x)\times[0,y)\biggr) \ , \label{4.15}
\end{equation} 

\noindent
which implies that $\mu_{\rm ph}$ shrinks as time goes on. This can be seen
immediately from the functional equation for $G$. For example, when $y\in
[0,\Lambda_1]$,

\begin{eqnarray}
\mu_{\rm ph}\Bigl(\Psi^{-1} \bigl\{ [0,x)\times [0,y) \bigr\} \Bigr) &=&
\mu_{\rm ph}\Bigl([0,l x)\times [0,y/\Lambda_1) \Bigr)
= l x G(y/\Lambda_1)  \nonumber \\
&=& (l+r) x G(y) = (l+r) \ \mu_{\rm ph}\Bigl(
[0,x)\times [0,y)\Bigr) \ . \nonumber
\end{eqnarray}

\noindent
Then, since the measure $\mu_{\rm ph}$ satisfies

\begin{eqnarray}
\mu_{\rm ph}\Bigl(E \Bigr) = {\displaystyle \mu_{\rm ph}\Bigl(\Psi^{-1} E 
\Bigr) \over \displaystyle \mu_{\rm ph}\Bigl(\Psi^{-1} [0,1)^2 \Bigr) }
\ , \nonumber
\end{eqnarray}

\noindent
for any Borel set $E \subset [0,1)^2$, it is conditionally invariant
\cite{20,20b}.


Now we turn to an invariant measure $\mu_{\rm in}$ on the repeller,
which is defined by

\begin{equation}
\mu_{\rm in}\Bigl([0,x)\times [0,y) \Bigr) = H(x) G(y)  \ . \label{4.16}
\end{equation}

\noindent
The invariance can be seen straightforwardly from the functional
equations for $G$ and $H$. For example, when $y\in [b,\Lambda_2+b]$, one
has

\begin{eqnarray}
\mu_{\rm in}\Bigl(\Psi^{-1} \bigl\{ [0,x)\times [0,y) \bigr\} \Bigr) &=& 
\mu_{\rm in}\Bigl([0,l x)\times [0,1) \cup [a, r x+a) \times 
[0,(y-b)/\Lambda_2) \Bigr) \nonumber \\
&=& \mu_{\rm in}\Bigl([0,l x)\times [0,1) \Bigr)+\mu_{\rm
in}\Bigl([a,r x+a)\times [0,(y-b)/\Lambda_2) \Bigr) \nonumber  \\
&=& H(l x) G(1) + \Bigl\{
H(r x+a)-H(a) \Bigr\} G\bigl((y-b)/\Lambda_2)\bigr) \nonumber \\
&=& H(x) {\displaystyle l \over \displaystyle l+r} +
{\displaystyle r\over \displaystyle l+r}
H(x) G\bigl((y-b)/\Lambda_2)\bigr) \nonumber \\ 
&=& H(x) G(y) =  \mu_{\rm in}\Bigl([0,x)\times [0,y) \Bigr) \ . \nonumber 
\end{eqnarray}

\noindent
Since $H(x)$ is a fractal function similar
to those discussed in Sec.II, the invariant measure is singular both along 
the contracting and expanding directions. This can be easily understood 
as follows: 
since the map $\Psi$ eventually transforms the unit square 
into the unstable manifold of the repeller, which is a direct product of a
Cantor set $C_1$ along $y$ and the unit interval  along $x$: $[0,1]\times C_1$,
the measure becomes singular along $y$.
On the other hand, only the orbits starting from the stable manifold of the
repeller remain in the unit square and the stable manifold is a direct product
set of the unit interval along $y$ and a Cantor set $C_2$ along $x$: $C_2\times
[0,1]$. 
Thus, the  invariant set is a subset of $C_2\times [0,1]$.  As a result, the
invariant measure becomes singular also along $x$. Actually, the direct product
of the two Cantor sets $C_2\times C_1$ is the fractal repeller of $\Psi$.


In a similar argument to the derivation of (\ref{4.2}), one can show that the
invariant measure $\mu_{\rm in}$ is mixing with respect to $\Psi$. 
As shown in Appendix C, the invariant measure $\mu_{\rm in}$ is a Gibbs measure
\cite{1,14,19b,20,20b,21,22,43,HOY}.

\vskip 0.7 truecm

\section{A BAKER-TYPE MAP UNDER A FLUX BOUNDARY CONDITION}


To illustrate a stationary state for an open system with a flux boundary
condition, we study a simple model of a chemical reaction. 
In simple reactions such as $R \leftrightarrow I \leftrightarrow P$, the
reactant $R$, the intermediate $I$ and the product $P$ consist of the same atoms,
and they can be specified by configurations of atoms, or points in the
atomic-configuration space. An example of the reaction $R \leftrightarrow I
\leftrightarrow P$ is an isomerization, or a change in the conformation of a
molecule, such as the ``chair'' to ``boat'' transformation of cyclohexane, where
$R$ is the chair-shaped isomer, $P$ is the boat-shaped isomer and $I$ is an
intermediate unstable isomer, all of which consist of six carbon atoms and twelve
hydrogen atoms. 
Hence, the reaction process can be regarded as a dynamical process where 
each trajectory starting from a reactant state to a product state
represents an individual reaction $R\to P$  (cf. Ref.
\cite{React} and references therein).
A connection between chemical reactions and the escape-rate formalism was
investigated by Dorfman and Gaspard \cite{19b} and a Baker-type model of a
chemical reaction was studied by Elskens and Kapral
\cite{ElKa}.
In this section, we introduce a Baker-type model of a reaction $R
\leftrightarrow I \leftrightarrow P$ and study its statistical properties
for two cases: In one case, the system is closed, and in the other case,
a flux boundary condition is imposed. Now we begin with the phenomenology.

\subsection{Phenomenology}

For a chemical reaction $R \leftrightarrow I \leftrightarrow P$, the
discrete-time version of the phenomenological rate equation is

\begin{eqnarray}
R_{t+1} &=& k_{RR} R_t + k_{RI} I_t \ , \nonumber \\
I_{t+1} &=& k_{IR} R_t + k_{II} I_t + k_{IP} P_t \ , \label{rateEQ} \\
P_{t+1} &=& k_{PI} I_t + k_{PP} P_t \ , \nonumber 
\end{eqnarray}

\noindent
where $R_t$, $I_t$, and $P_t$ are concentrations of the reactant $R$,
intermediate $I$, and product $P$, respectively, and $k_{AB}$ ($A, B = R, I,$ or
$P$) are rate coefficients. Since the sum $R_t + I_t + P_t$ is preserved,
the rate coefficients satisfy a sum rule :

\begin{eqnarray}
k_{RR}+k_{IR}&=&1 \ , \nonumber \\
k_{RI}+k_{II}+k_{PI}&=&1 \ , \label{sum} \\
k_{IP} + k_{PP}&=&1 \ . \nonumber 
\end{eqnarray}

\noindent
The stationary state solution of Eq. (\ref{rateEQ}) is then given by

\begin{equation}
{\displaystyle R_{\rm st} \over \displaystyle I_{\rm st} } = {\displaystyle
k_{RI}
\over \displaystyle k_{IR}} \ , \qquad \qquad 
{\displaystyle P_{\rm st} \over \displaystyle I_{\rm st} } = {\displaystyle
k_{PI}
\over \displaystyle k_{IP}} \ . \label{ST} 
\end{equation}

Now we consider a stationary solution of Eq. (\ref{rateEQ}) under a flux boundary
condition, where the concentrations of the reactant $R$ and the product $P$ are
fixed to given values $R_{\rm ex}$ and $P_{\rm ex}$, respectively.
This may be realized e.g., by introducing source terms to the equations for the
reactant and product and by adjusting them so as to keep the values of $R_t$
and $P_t$ constant.  
Eq. (\ref{rateEQ}) is then reduced to

\begin{equation}
I_{t+1} = k_{IR} R_{\rm ex} + k_{II} I_t + k_{IP} P_{\rm ex} \ , \label{rateEQ2} 
\end{equation}

\noindent
which has a stationary solution

\begin{equation}
I_{\rm fl} = {\displaystyle k_{IR} R_{\rm ex} + k_{IP} P_{\rm ex} \over
\displaystyle 1- k_{II}} \ . \label{fluxST}
\end{equation}

\noindent
The deviation $\delta I_t \equiv I_t - I_{\rm fl}$ of the intermediate
concentration from the stationary value decays exponentially :

\begin{equation}
\delta I_t = k_{II} \ \delta I_{t-1} = \cdots = k_{II}^t \ \delta I_0 \ .
\label{decay}
\end{equation}

\noindent
The stationary state admits a non-vanishing concentration flow from
the reactant to the product :

\begin{equation}
J_{R\to P} = k_{PI} I_{\rm fl} - k_{IP} P_{\rm ex} =
{\displaystyle k_{PI} k_{IR} \over \displaystyle 1-k_{II}} R_{\rm ex}
- {\displaystyle k_{RI} k_{IP} \over \displaystyle 1-k_{II}} P_{\rm ex} \ ,
\label{flowPH}
\end{equation}

\noindent
and, hence, can be regarded as a stationary state under a flux boundary
condition.

\subsection{A Baker-type model and a stationary state for a closed system}

We introduce a Baker-type model of the reaction process $R \leftrightarrow I
\leftrightarrow P$. Microscopic dynamical states of each species $R$, $I$, or
$P$ are represented by the points in a unit square $(0,1]^2$.  Area preserving
asymmetric Baker maps are used to describe the dynamics of the reactant and
product states, and a Baker map similar to the one discussed in the previous
section is used for the intermediate state dynamics.
The model is then defined by (Fig.~\ref{figFL}):

\begin{eqnarray}
\Psi'(I:x,y) &=& \cases{ \left(I: {\displaystyle x \over \displaystyle l},
\Lambda_1 y
\right) \ , &$x\in (0,l]$
\cr 
\left(R: {\displaystyle x-l \over \displaystyle a-l}, (b-\Lambda_1) y \right) \
, &$x\in (l,a]$
\cr 
\left(I: {\displaystyle x-a \over \displaystyle r}, \Lambda_2 y+ b \right) \ ,
&$x\in (a,a+r]$
\cr
\left(P: {\displaystyle x-a-r \over \displaystyle 1-a-r}, (1-b-\Lambda_2) y
\right) \ , &$x\in (a+r,1]$ \cr}
\label{Imap} \\ \nonumber \\ \nonumber \\
\Psi'(R:x,y) &=& \cases{ \left(I: {\displaystyle x \over
\displaystyle b-\Lambda_1}, (b-\Lambda_1) y+\Lambda_1 \right)
\ , &$x\in (0,b-\Lambda_1]$ \cr \cr
\left(R: {\displaystyle x-b+\Lambda_1 \over \displaystyle 1-b+\Lambda_1},
(1-b+\Lambda_1) y+b-\Lambda_1 \right) \ , &$x\in (b-\Lambda_1,1]$ \cr}
\label{Rmap} \\ \nonumber \\ \nonumber \\
\Psi'(P:x,y) &=& \cases{ \left(I: {\displaystyle x \over
\displaystyle 1-b-\Lambda_2}, (1-b-\Lambda_2) y+b+\Lambda_2
\right) \ , &$x\in (0,1-b-\Lambda_2]$ \cr \cr
\left(P: {\displaystyle x-1+b+\Lambda_2 \over \displaystyle b+\Lambda_2},
(b+\Lambda_2) y+1-b-\Lambda_2 \right)
\ , &$x\in (1-b-\Lambda_2,1]$ \cr}
\label{Pmap}
\end{eqnarray}

\noindent
where $(x,y)$ denotes a point in a unit square and $\alpha$ ($\alpha=R, I$, or
$P$) distinguishes different species.

As before, we study the time evolution of a measure starting from an initial
measure which is absolutely continuous with respect to the two-dimensional
Lebesgue measure. The Frobenius-Perron equation for the density function
$\rho_t(\alpha:x,y)$  ($\alpha=R, I$, or $P$) is obtained from

$$
\rho_{t+1}(\alpha:x,y) = \sum_{\beta=R,I,P} \int_{[0,1]^2} dx dy
\delta\left((\alpha:x,y)-\Psi'(\beta:x',y')\right) \rho_t(\beta:x',y') \ ,
$$

\noindent
where the delta function $\delta\left((\alpha:x,y)-(\beta:x',y')\right)$ stands
for the product $\delta_{\alpha,\beta} \delta(x-x') \delta(y-y')$. 
By integrating the Frobenius-Perron equation with respect to $y$, one obtains the
evolution equation for the partially integrated distribution function
$Q_t(\alpha:x,y)=\int_0^y dy' \rho_t(\alpha:x,y')$ :

\begin{equation}
Q_{t+1}(\alpha:x,y) = {\bar {\cal F}}Q_t(\alpha:x,y) + {\bar R}_t(\alpha:x,y)
\ , \label{EvEQ1}
\end{equation}

\noindent
where a contraction mapping ${\bar {\cal F}}$ and a functional ${\bar R}_t$ of 
$Q_t(\alpha:x,1)$ are given in Appendix D (cf. Eqs. (\ref{Imeas}), (\ref{Rmeas}),
(\ref{Pmeas}) and  Eqs. (\ref{Iresi}), (\ref{Rresi}), (\ref{Presi}),
respectively).

Now we consider the case where the initial measure is uniform along the
expanding $x$ direction and thus, $Q_0(\alpha:x,y)$ does not depend on $x$.
Then, as seen from the expressions of ${\bar {\cal F}}$ and ${\bar
R}_t$, the partially integrated function $Q_t(\alpha:x,y)$ at time $t$ is also
independent of $x$. Particularly, its value $Q_t(\alpha:x,y=1)\equiv Q_t(\alpha)$
at $y=1$ obeys

\begin{eqnarray}
Q_{t+1}(R) &=& (1-b+\Lambda_1) Q_t(R) + (a-l) Q_t(I) \ , \nonumber \\
Q_{t+1}(I) &=& (b-\Lambda_1) Q_t(R) + (l+r) Q_t(I) + (1-b-\Lambda_2) Q_t(P) \ ,
\label{rateEQmap}
\\ 
Q_{t+1}(P) &=& (1-a-r) Q_t(I) + (b+\Lambda_2) Q_t(P) \ . \nonumber 
\end{eqnarray}

\noindent
Since the distribution is uniform along the $x$ direction,
$Q_t(\alpha)$ is equal to the total probability of finding the system in a
species $\alpha$ : $Q_t(\alpha) = \int dx dy \rho_t(\alpha:x,y)$.
Therefore, Eq. (\ref{rateEQmap}) has exactly the same form as Eq. (\ref{rateEQ})
where the corresponding rate coefficients are given by

\begin{eqnarray}
k_{RR}&=&1-b+\Lambda_1 \ , \qquad k_{RI}=a-l \ , \nonumber  \\
k_{IR}&=&b-\Lambda_1  \ , \qquad \qquad k_{II}=l+r \ ,  \qquad
k_{IP}=1-b-\Lambda_2 \ , \label{Coef}\\ 
k_{PI}&=&1-a-r \ , \qquad k_{PP}=b+\Lambda_2 \ .
\nonumber
\end{eqnarray}

\noindent
Note that the rate coefficients given above satisfy the sum rule Eq. (\ref{sum})
and hence, the sum $Q_t(R) + Q_t(I) + Q_t(P)$ is constant in time. We also
remark that the rate coefficients (\ref{Coef}) admit a simple geometrical
interpretation. As an example, we consider $k_{RI}$, which is the transition
probability from the intermediate $I$ to the reactant $R$. According to the
definition of the map $\Psi'$, a rectangle $(l,a]\times (0,1]$ moves from the
intermediate states to the reactant states and, thus, its Lebesgue area $(a-l)$
corresponds to the transition probability $k_{RI}$ from $I$ to $R$, or we have
the second equation of (\ref{Coef}). The other rate coefficients can be obtained
in the same way.

According to the discussion given in the previous subsection, the stationary
solution of (\ref{rateEQmap}) is

\begin{equation}
Q_{\rm st}(R)={a-l \over b-\Lambda_1} Q_{\rm st}(I) \ , \qquad
Q_{\rm st}(P)={1-a-r \over 1-b-\Lambda_2} Q_{\rm st}(I) \ , \label{STmap}
\end{equation}

\noindent
and the corresponding distribution is given by

\begin{eqnarray}
Q_{\rm st}(R:x,y) &=& {a-l\over b-\Lambda_1}Q_{\rm st}(I) q_{\rm st}(R:y) \ ,
\qquad 
Q_{\rm st}(I:x,y) = Q_{\rm st}(I) q_{\rm st}(I:y) \ , \nonumber \\
Q_{\rm st}(P:x,y) &=& {1-a-r\over 1-b-\Lambda_2}Q_{\rm st}(I) q_{\rm st}(P:y)
\ , \label{STmapDis1}
\end{eqnarray}

\noindent
where the singular functions $q_{\rm st}(\alpha:y)$ ($\alpha=R,I,$ and $P$) are
the unique solutions of de Rham equations

\FL
\begin{eqnarray}
q_{\rm st}(I:y) &=& \cases{ l \ q_{\rm st}\left(I: {\displaystyle y \over
\displaystyle \Lambda_1}
\right) \ , &$y\in (0,\Lambda_1]$
\cr 
l+(a-l) q_{\rm st}\left(R: {\displaystyle y-\Lambda_1 \over
\displaystyle b-\Lambda_1} \right) \ , &$y\in (\Lambda_1,b]$
\cr 
a+r \ q_{\rm st}\left(I: {\displaystyle y-b \over \displaystyle
\Lambda_2}\right)
\ , &$y\in (b,\Lambda_2+b]$
\cr
a+r+(1-a-r) q_{\rm st}\left(P: {\displaystyle y-b-\Lambda_2
\over \displaystyle 1-b-\Lambda_2} \right) \ , &$y\in (\Lambda_2+b,1]$ \cr}
\label{STmapDis2}  \\ \nonumber \\ \nonumber \\
q_{\rm st}(R:y) &=& \cases{ (b-\Lambda_1) q_{\rm st}\left(I: {\displaystyle
y \over \displaystyle b-\Lambda_1} \right)
\ , &$y\in (0,b-\Lambda_1]$ \cr 
b-\Lambda_1+(1-b+\Lambda_1) q_{\rm st}\left(R: {\displaystyle y-b+\Lambda_1
\over \displaystyle 1-b+\Lambda_1}\right) \ , &$y\in (b-\Lambda_1,1]$ \cr}
\label{STmapDis3}  \\ \nonumber \\ \nonumber \\
q_{\rm st}(P:y) &=& \cases{ (1-b-\Lambda_2) q_{\rm st}\left(I: {\displaystyle y
\over \displaystyle 1-b-\Lambda_2}\right) \ , &$y\in (0,1-b-\Lambda_2]$ \cr
1-b-\Lambda_2+(b+\Lambda_2) q_{\rm st}\left(P: {\displaystyle y-1+b+\Lambda_2
\over \displaystyle b+\Lambda_2} \right)
\ . &$y\in (1-b-\Lambda_2,1]$ \cr}
\label{STmapDis4}
\end{eqnarray}

\noindent
One can show, by exactly the same method as before, that the partially
integrated function $Q_t(\alpha:x,y)$ approaches the stationary state $Q_{\rm
st}(\alpha:x,y)$ provided that the initial function $Q_0(\alpha:x,y)$ is
continuously differentiable with respect to $x$. Then since the asymptotic
measure $Q_{\rm st}(\alpha:x,y)$ is absolutely continuous with respect to the
Lebesgue measure along the expanding $x$ direction, it it the SRB measure. Note
that, in this case, the quantity $Q_{\rm st}(I)$ is a functional of the initial
distribution $Q_0(\alpha:x,y)$.

\subsection{A stationary state under a flux boundary condition}

As shown in Sec. VA, one has a flux from the reactant to the product when
their concentrations are fixed to the values different from the equilibrium
ones. This can be realized in the Baker-type model by fixing the measures of the
unit squares corresponding to the reactant and the product to uniform
Lebesgue measures with different densities.
So we set $Q_t(R:x,y)=R_{\rm
ex} y$ and $Q_t(P:x,y)=P_{\rm ex} y$ ($t=0,1,\cdots$). 
Note that the same
procedure was used to achieve the flux boundary condition for the finite
multi-Baker chain \cite{15}.
Then, we have only one time-dependent variable
$Q_t(I:x,y)$, which will be abbreviated as $Q_t(I:x,y)\equiv Q_t(x,y)$.
Then  the equation of motion (\ref{EvEQ1}) for the partially 
integrated distribution function $Q_t$ becomes

\begin{equation}
Q_{t+1}(x,y) = {\cal F}' Q_t(x,y)+R_t(x,y)+S(y) \ , \label{5.1a}
\end{equation}

\noindent
where the contraction mapping ${\cal F}'$ and $Q_t(x,1)$-dependent part
$R_t(x,y)$ are given by Eqs. (\ref{4.4b}) and (\ref{4.4c}) of Appendix B,
respectively, and the source term $S(y)$ is due to the reactant and product
states

\begin{equation}
S(y) = \cases{ 0 \ ,
&$y\in [0,\Lambda_1]$ \cr 
R_{\rm ex} (y-\Lambda_1) \ ,
&$y\in (\Lambda_1, b)$ \cr 
R_{\rm ex} (b-\Lambda_1)  \
, &$y\in [b,\Lambda_2+b]$ \cr  
R_{\rm ex} (b-\Lambda_1) + P_{\rm ex} (y-b-\Lambda_2) \ .
&$y\in (\Lambda_2+ b,1]$ \cr 
} \label{5.1b}
\end{equation}


\noindent
By exactly the same argument as before, Eqs. 
(\ref{5.1a}) and (\ref{5.1b}) is found to have a solution

\begin{eqnarray}
Q_t(x,y) &=& {b-\Lambda_1 \over 1-l-r} \ R_{\rm ex} \ \eta_R(y) +
{1-b-\Lambda_2 \over 1-l-r} \ P_{\rm ex} \ \eta_P(y) \nonumber \\
& & + \nu^t \int_0^1 dH(x')
\biggl\{Q_0(x',1) -{b-\Lambda_1 \over 1-l-r} \ R_{\rm ex}  -
{1-b-\Lambda_2 \over 1-l-r} \ P_{\rm ex} \biggr\} G(y) + {\rm
O}(t\lambda'^t) \ , \label{5.FLUX}
\end{eqnarray}

\noindent
where 
$G(y)$ and $H(x)$ are singular functions introduced in Sec. IV (cf. Eqs.
(\ref{4.10b}) and (\ref{4.7})), and 
the functions $\eta_R(y)$ and $\eta_P(y)$
are the unique solutions of de Rham equations

\begin{eqnarray}
\eta_R(y) &=& \cases{ l \ \eta_R\bigl({\displaystyle y \over \displaystyle
\Lambda_1} \bigr) \ , &$y\in [0,\Lambda_1]$ \cr  
1-r + {\displaystyle 1-l-r \over \displaystyle
b-\Lambda_1} (y-b) \ , &$y\in (\Lambda_1, b)$ \cr 
r \ \eta_R\bigl({\displaystyle y-b
\over  \displaystyle \Lambda_2}\bigr) + 1-r \ , &$y\in
[b,\Lambda_2+b]$ \cr  
1 \ , &$y\in
(\Lambda_2+b,1]$ \cr 
} \label{q_inf1} \\ \nonumber \\ \nonumber \\
\eta_P(y) &=& \cases{ l \ \eta_P\bigl({\displaystyle y \over \displaystyle
\Lambda_1} \bigr) \ , &$y\in [0,\Lambda_1]$ \cr  
l \ , &$y\in (\Lambda_1, b)$ \cr 
r \ \eta_P\bigl({\displaystyle y-b
\over  \displaystyle \Lambda_2}\bigr) + l \ , &$y\in
[b,\Lambda_2+b]$ \cr  
1 + {\displaystyle
1-l-r\over \displaystyle 1-b-\Lambda_2}(y-1) 
\ . &$y\in
(\Lambda_2+b,1]$ \cr 
} \label{q_inf2} 
\end{eqnarray}

\noindent
Therefore the stationary measure 
$\mu_{\rm fl}$ \ for this open system is given by

\begin{equation}
\mu_{\rm fl} \biggl([0,x)\times [0,y)\biggr)  =  {b-\Lambda_1 \over 1-l-r} \
R_{\rm ex} \ x \ \eta_R(y) + {1-b-\Lambda_2 \over 1-l-r} \ P_{\rm ex} \
x \ \eta_P(y) 
\ . \label{mu_fl}
\end{equation}

\noindent
As it is smooth (i.e., absolutely continuous 
with respect to the Lebesgue
measure) along 
the expanding coordinate $x$, it is an SRB measure.
However, the measure
$\mu_{\rm fl}$ is different from the SRB measures 
obtained before since it is absolutely continuous
with respect to $y$. It is only weakly singular in the sense that the density is
discontinuous on a Cantor-like set 
(cf. Fig.~\ref{fig9}). Actually, such stationary measures
become truely singular only for infinite systems \cite{15,17}.
It is interesting to observe that the conditionally invariant measure $\mu_{\rm
ph} \biggl([0,x)\times [0,y)\biggr) = x G(y)$ appears in the time evolution of
the measure (\ref{5.FLUX}) as the decay mode, i.e., as the Pollicott-Ruelle
resonance.

Now we investigate macroscopic aspects. The total measure $\mu_t
\biggl((0,1]^2 \biggr)\equiv \int_0^1 dx Q_t(x,1)$ correponding to the
concentration of the intermediate $I$ is

\begin{equation}
\mu_t \biggl((0,1]^2 \biggr) = {b-\Lambda_1 \over 1-l-r} \ R_{\rm ex} \
 + {1-b-\Lambda_2 \over 1-l-r} \ P_{\rm ex} + \nu^t \delta \mu_0  
+{\rm
O}(t\lambda'^t) \ , \label{5.FLUX2}
\end{equation}

\noindent
where $\delta \mu_0$ stands for the deviation of the initial measure from the
stationary state $\mu_{\rm fl}$

$$
\delta \mu_0 \equiv \int_0^1 dH(x')
\biggl\{Q_0(x',1) -{b-\Lambda_1 \over 1-l-r} \ R_{\rm ex}  -
{1-b-\Lambda_2 \over 1-l-r} \ P_{\rm ex} \biggr\} \ .
$$

\noindent
The stationary state $\mu_{\rm fl}$ admits a flux $J_{R\to P}$ from the reactant
state to the product state.
Indeed, the measure $\mu_{\rm fl}\Bigl(
(a+r,1]\times (0,1]\Bigr)$ is transfered to the product state and the measure
$(1-b-\Lambda_2) P_{\rm ex}$ is transfered from there. This implies the
existence of the flux $J_{R\to P}$ given by

\begin{eqnarray}
J_{R\to P} &=& \mu_{\rm fl}\Bigl( (a+r,1]\times (0,1]\Bigr) - (1-b-\Lambda_2)
P_{\rm ex} \nonumber \\
&=& {(1-a-r)(b-\Lambda_1) \over 1-l-r} R_{\rm ex} - 
{(a-l)(1-b-\Lambda_2) \over 1-l-r} P_{\rm ex} \ . \label{5.flow}
\end{eqnarray}

\noindent
With the specification (\ref{Coef}) of the rate coefficients, the expressions 
(\ref{5.FLUX2}) of the stationary measure and the decay mode as well as 
(\ref{5.flow}) of the flux agree with their phenomenological counterparts
(\ref{fluxST}), (\ref{decay}) and (\ref{flowPH}), respectively.
In short, the measure $\mu_{\rm fl}$ describes a nonequlibrium stationary state
which has a non-vanishing flux $J_{R\to P}$.


\vskip 1truecm

\section{CONCLUSIONS}

We have explicitly constructed SRB measures for three 
Baker-type maps employing the similarlity between the de Rham equation 
and the evolution equation of partially integrated distribution functions.  
Now we give a few more remarks.

\vskip .7 truecm

\noindent a) \
In our examples discussed in Secs. III and IV, we have encountered different
types of contracting dynamics that lead to invariant states all of which
have singular measures supported on fractal sets. 
In the first case we
considered a non-conservative reversible Baker map on the unit square which
globally preserves the area of the square but forms an attractor-repeller pair
due to the local contraction and expansion properties of the map. 
This attractor is fractal since it has the information dimension less than $2$,
but it is dense in the unit square. 
In the second example of the Baker map with a Cantor-like invariant set, we
considered two different cases - global contraction onto a fractal set with and
without escape of points from the unit square. When there is no escape, the
invariant set is a fractal attractor which is a non-dense subset of the unit
square.
If we add the possibility of escape, then the invariant set
is fractal in both
$x$ and $y$ directions. 

\vskip 0.7 truecm

\noindent b) \
It is worth mentioning that one can convert a physical measure for a system
with escape to the proper invariant measure on the repeller by incorporating
into the averaging process described in Eq. (\ref{1.5}) both the characteristic
function on the region of the phase space from which points escape, as well
as a ``survival'' function which is unity if the phase point is in the region
for the time interval $0<\tau<T$, with $T>t$, and zero otherwise. Then by
taking the limit $T\to \infty$, one recovers the invariant measure on the
repeller. Such a procedure was used by van Beijeren and Dorfman in order to
correctly compute the Lyapunov exponents on the repeller for a Lorentz gas 
\cite{43}. This procedure is closely related to one described by Hunt, Ott, and
Yorke \cite{HOY} to obtain natural measures on invariant sets.

\vskip 0.7 truecm


\noindent c) \ 
We have encountered three different physical measures, which are characterized
by the smoothness along the expanding direction. For closed non-conservative
systems, the physical measures are singular and invariant. For an
open system under an absorbing boundary condition, the physical measure is
singular and conditionally invariant. And for an open system under a flux
boundary condition, the physical measure is invariant, but absolutely continuous
with respect to the two-dimensional Lebesgue measure. It is only weakly singular
in the sense that the density is discontinuous on a Cantor-like set. 

It is interesting to note that the conditionally invariant measures appear as the
decay modes (i.e., the Pollicott-Ruelle resonances) for an open system under a
flux boundary condition. 
Such a relation exists more generally. The Pollicott-Ruelle resonances are
defined as the functionals $\Gamma$ acting on a dynamical variable $\varphi$
\cite{16,Po,Ru,PoRuRes,29,36,36b,37}, 
which satisfy, in case of a map,

$$
\bra \Gamma , \varphi\circ \Phi^t \ket = \zeta^t \ \bra \Gamma , \varphi \ket
\ ,
$$

\noindent
where $\Phi$ is a map and $\zeta$ ($|\zeta|<1$) is a decay rate. When the
characteristic function $\chi_A$ of any Borel set $A$ is in the domain of the
functional
$\Gamma$ and $\bra \Gamma , \chi_M \ket\not=0$ with $M$ the whole phase space,
then the (possibly complex) measure defined by

$$
\mu(A) = \bra \Gamma , \chi_A \ket \ , 
$$

\noindent
is conditionally invariant since $\mu(\Phi^{-1}A)/\mu(\Phi^{-1}M) = \mu(A)$.
Such examples are the hydrodynamic modes for open Hamiltonian systems
\cite{16,36,36b}. 

\vskip 0.7 truecm

\noindent d) \ 
For a non-conservative reversible map $\Phi$, we find different SRB
measures ${\mu}_{\rm ph}$ and ${\bar \mu}_{\rm ph}$ for the forward and backward
time evolutions, respectively. For each time evolution, one plays a role of
an attracting measure and the other of a repelling measure. 
This is a key element of the compatibility between dynamical
reversibility and irreversible behavior of statistical ensembles. Indeed, when
the dynamics is reversible and statistical ensembles irreversibly approach a
stationary ensemble ${\mu}_{\rm ph}$ for $t\to +\infty$, there should exist
another stationary ensemble ${\bar \mu}_{\rm ph}$ which is obtained from
${\mu}_{\rm ph}$ by the time reversal operation. Since the new stationary
ensemble ${\bar \mu}_{\rm ph}$ is repelling, its existence is compatible with
the irreversible behavior of statistical ensembles. 
A similar situation was observed for 
open Hamiltonian systems under a flux boundary condition \cite{15,16}.
Such systems admit two different stationary states, one is the time 
reversed state of the other.  In this case, an attracting stationary state for
the forward time evolution obeys Fick's law and a repelling state obeys
anti-Fick's law.

\vskip 0.7 truecm

\noindent e) \ 
As mentioned in Introduction, an SRB measure for a map may be constructed as
follows \cite{Ott}: 1) \ Approximate
the measure by iterating an initial constant density finite times, 2) \
calculate the average with its result and 3) \ take the limit of infinite
iterations.  On the other hand, in our construction of an SRB measure, a
functional equation for the partially integrated distribution function similar
to the de Rham equation is derived from the evolution equation for measures. 
Note that the functional equation is a direct consequence of the self-similarity
of the measure. 
An iterative method of solving the functional equation is similar to the
procedure explained above. 
However, our method has some
advantages. Firstly, the de Rham-type functional equations are fixed
point equations for contraction mappings and, as a result, the iterative solution
strongly converges to the limit. On the contrary, the iterative approximation of
the density function does not converge by itself. Therefore, one can obtain a
good approximation to the cumulative distribution function of the SRB measure by
less iterations than by the conventional method. Secondly, in the functional
equation approach, one can systematically extract exponentially decaying terms
which are typically higher order derivatives of singular functions in the sense
of Schwartz' distributions. Thirdly, although the measure is defined as the
solution of a functional equation, the average values of certain dynamical
functions can be calculated analytically as illustrated in Sec. II.
So far, the functional equation method was mainly applied to piecewise linear
one-dimensional maps and Baker-type maps \cite{15,36b,37}, but we believe
that the method can also be applied to other systems if the expanding and
contracting directions can be well-separated.


\acknowledgments

ST and JRD thank Prof. T. T\'el and 
Prof. I. Kondor for having invited them to an
exciting Summer School/Workshop: ``Chaos and Irreversibility'' (Etov\"os
University, Budapest, 31 August - 6 September, 1997) and for their warm
hospitality. The authors are grateful to Prof. P. Gaspard for fruitful 
discussions and helpful comments, 
particularly on the relation between the
physical measure and the conditionally invariant measure, 
and to Profs. T. T\'el,
J. Vollmer and W.  Breymann for fruitful discussions, particularly on their work
about multi-Baker maps. This  work is a part of the project of Institute for
Fundamental Chemistry, supported by Japan Society for the Promotion of Science -
Research for the Future Program (JSPS-RFTF96P00206). Also this work is partially
supported by a Grant-in-Aid for Scientific Research and a grant under the
International Scientific Research Program both from Ministry of Education,
Science and Culture of Japan. JRD also wishes to thank Prof. Henk van Beijeren,
and Dr. Arnulf Latz for many helpful discussions as well as the National Science
Foundation for support under grant PHY-96-00428.

\vskip 0.7 truecm

\appendix

\section{Proofs of Eq. (\ref{3.5}) and the lemma}

\subsection{Derivation of Eq. (\ref{3.5})}

As explained in Sec. III, the evolution equation (\ref{3.4}) for the partially
averaged distribution function $P_t(x,1)\equiv \int_0^1 dy' \rho_t(x,y')$ 

\begin{equation}
P_{t+1}(x,1)= r \ P_t\bigl(r \ x+l,1\bigr) + l \  P_t\bigl(l \ x,1\bigr) \ , 
\label{A1}
\end{equation}

\noindent
is the Frobenius-Perron equation for a one-dimensional exact map $Sx=x/l$ (for
$x\in [0,l]$) and $Sx=(x-l)/r$ (for $x\in (l,1]$), which admits the Lebesgue
measure as the invariant measure.  As a consequence,
the integral of $P_t(x,1)$ over the unit interval [0,1] is invariant. Indeed, by
integrating Eq. (\ref{A1}), we have 

$$
\int_0^1 dx P_{t+1}(x,1)=\int_0^1 dx P_t(x,1)=\cdots =\int_0^1 dx P_0(x,1) \ .
$$

\noindent
On the other hand, since $\rho_0$ and, hence, $P_0$ is continuously
differentiable in $x$, Eq. (\ref{A1}) implies that $P_t(x,1)$ ($t=1,2,\cdots$)
are also continuously differentiable in $x$. And the derivative $\partial_x
P_t(x,1)$ satisfies the equation

$$
\partial_x P_{t+1}(x,1)= r^2 \ \partial_x P_t\bigl(r \ x+l,1\bigr) 
+ l^2 \
\partial_x P_t\bigl(l \ x,1\bigr) \ , 
$$

\noindent
which leads to an inequality

$$
{ \Norm{\partial_x P_{t+1}(\cdot,1)} }\le (r^2+l^2) \ { \Norm{\partial_x
P_t(\cdot,1)} } \le \lambda { \Norm{\partial_x
P_t(\cdot,1)} } \le \cdots \le \lambda^{t+1} { \Norm{\partial_x
P_0(\cdot,1)} }\ , 
$$

\noindent
where the function norm is defined by ${ \Norm{\partial_x P_t(\cdot,1)} }
\equiv \sup_{x\in[0,1]} |\partial_x P_t(x,1)|$ and $\lambda=\max(l,\ r)<1$.
Since the function $P_t(x,1)$ can be represented as

$$
P_t(x,1)-\int_0^1 dx' P_t(x',1) = \int_0^1 dx' \ x' \ \partial_{x'} P_t(x',1)
- \int_x^1 dx' \partial_{x'} P_t(x',1) \ , 
$$

\noindent
we finally obtain

\begin{eqnarray}
\big| P_t(x,1)-\int_0^1 dx' P_0(x',1) \big| &\le& \Bigl\{ \int_0^1 dx' \ x' +
\int_x^1 dx'\Bigr\}  {\Norm{\partial_x P_t(\cdot,1) } } \nonumber \\
&\le& {3\over 2}  {\Norm{\partial_x P_t(\cdot,1) } }
\le {3\over 2}  \lambda^t {\Norm{\partial_x P_0(\cdot,1) } } \ , \nonumber
\end{eqnarray}

\noindent
or Eq. (\ref{3.5}).

\subsection{Proof of the lemma}

Since the statement of the lemma given in the text is not technically complete,
we first give the precise statement and then prove it.

\vskip 10 pt

\noindent{\bf Lemma} \ \ \ {\it
Let ${\cal F}$ be a linear contraction mapping with the contraction constant
$0<\lambda<1$ defined on a (Banach) space of bounded functions equipped with the
supremum norm ${ \Norm{ \ \cdot \ } }$ (i.e., ${ \Norm{{\cal F} f} }\le
\lambda { \Norm{f} }$). And let $g_t\equiv g^{(0)}+\nu^t g^{(1)} 
+g^{(2)}_t$ be a
given function where $g^{(0)}$, $g^{(1)}$ and $g^{(2)}_t$ are bounded functions,
$\nu$ is a constant satisfying $\lambda<|\nu|\le 1$, 
and ${ \Norm{g^{(2)}_t} }\le
K \lambda^t$ with some constant $K>0$. Then, the solution of the functional
equation
\begin{equation}
f_{t+1} = {\cal F} f_t + g_t \ , \label{A3.6}
\end{equation}
is given by
\begin{equation}
f_t = f_\infty^{(0)} + \nu^t f_\infty^{(1)} + {\rm O}(t \lambda^t) \ ,
\label{A3.7a}
\end{equation}
where $f_\infty^{(0)}$ and $f_\infty^{(1)}$ are the unique solutions of the
following fixed point equations
\begin{eqnarray}
f_\infty^{(0)} &=& {\cal F} f_\infty^{(0)} + g^{(0)} \ , \label{A3.7b} \\
f_\infty^{(1)} &=& {1\over \nu}{\cal F} f_\infty^{(1)} + {g^{(1)}\over \nu} \ .
\label{A3.7c}
\end{eqnarray}
}

\vskip 8 pt

The proof of the lemma is straightforward: From Eq. (\ref{A3.6}), one obtains

$$
f_t = {\cal F}^t f_0+\sum_{s=0}^{t-1} {\cal F}^s g_{t-1-s} \ . 
$$

\noindent
By rewriting the right hand side in terms of $g^{(0)}$, $g^{(1)}$ and
$g^{(2)}_t$, this relation leads to

\begin{eqnarray}
f_t &=& \sum_{s=0}^\infty {\cal F}^s g^{(0)}+\nu^t  \sum_{s=0}^\infty 
\nu^{-s-1} {\cal F}^s g^{(1)} \nonumber \\
& &+ \biggl\{ {\cal F}^t f_0 +\sum_{s=0}^{t-1} {\cal
F}^s g^{(2)}_{t-1-s} -\sum_{s=t}^\infty {\cal F}^s g^{(0)}-\nu^t 
\sum_{s=t}^\infty \nu^{-s-1} {\cal F}^s g^{(1)} \biggr\} 
 \ . \nonumber 
\end{eqnarray}

\noindent
The first sum $\sum_{s=0}^\infty {\cal F}^s g^{(0)}$ in the right-hand side is
$f_\infty^{(0)}$ and the second sum  $\sum_{s=0}^\infty \nu^{-s-1} {\cal F}^s
g^{(1)}$ is $f_\infty^{(1)}$, which satisfy Eqs. \ (\ref{A3.7b}) and
(\ref{A3.7c}) respectively.  By repeatedly using the property of $\cal F$, we
have

$$
{ \Norm{f_t-f_\infty^{(0)}-\nu^t f_\infty^{(1)}} }
\le \lambda^t \biggl\{ { \Norm{f_0} } + {{ \Norm{g^{(0)} } } \over 1-\lambda } 
+ { { \Norm{g^{(1)} } } \over |\nu|-\lambda} + K \ t \biggr\} \ ,
$$

\noindent
which is O($t \lambda^t)$ and implies the desired result (\ref{A3.7a}).

\section{PHYSICAL MEASURE FOR A BAKER MAP WITH \\ A CANTOR-LIKE INVARIANT SET}

In this Appendix, we outline the derivation of Eq. (\ref{4.2}), which is quite
similar to that of Eq.(\ref{3.10}).

>From the definition (\ref{4.1}) of the map $\Psi$, one finds that 
the Frobenius-Perron equation governing the time evolution of the distribution
function $\rho_t(x,y)$ is given by

\begin{equation}
\rho_{t+1}(x,y) = \cases{ {\displaystyle l\over \displaystyle \Lambda_1}
\rho_t\bigl(l x, {\displaystyle y \over \displaystyle \Lambda_1} \bigr) \ ,
&$y\in [0,\Lambda_1]$ \cr 
0 \ ,
&$y\in (\Lambda_1, b)$ \cr 
{\displaystyle r\over \displaystyle \Lambda_2}
\rho_t\bigl(r x+a, {\displaystyle y- b \over \displaystyle \Lambda_2} 
\bigr) \
, &$y\in [b,\Lambda_2+b]$ \cr  0 \ ,
&$y\in (\Lambda_2+b,1]$ \cr 
} \label{4.3} 
\end{equation}

\noindent
which leads to a contractive time evolution of the partially integrated
distribution function  $Q_{t}(x,y)\equiv \int_0^y dy' \rho_t(x, y')$:

\begin{equation}
Q_{t+1}(x,y) = {\cal F}' Q_t(x,y)+R_t(x,y) \ , \label{4.4a}
\end{equation}

\noindent
where ${\cal F}'$ stands for a contraction mapping:

\begin{equation}
{\cal F}' Q_t(x,y) \equiv \cases{ l Q_t\bigl(l x, {\displaystyle y 
\over \displaystyle
\Lambda_1}, \bigr) \ , &$y\in [0,\Lambda_1]$ \cr 
0 \ ,
&$y\in (\Lambda_1, b)$ \cr 
r Q_t\bigl(r x+a, {\displaystyle y- b \over \displaystyle \Lambda_2} 
\bigr) \ , &$y\in [b,\Lambda_2+b]$ \cr  0 \
, &$y\in (\Lambda_2+b,1]$ \cr 
} \label{4.4b} 
\end{equation}

\noindent
and $R_t(x,y)$ is a function of $Q_t(x,1)$:

\begin{equation}
R_t(x,y) = \cases{ 0 \ ,
&$y\in [0,\Lambda_1]$ \cr 
l Q_t\bigl( l x, 1 \bigr) \ ,
&$y\in (\Lambda_1, b)$ \cr 
l Q_t\bigl( l x, 1 \bigr)  \
, &$y\in [b,\Lambda_2+b]$ \cr  
l Q_t\bigl( l x, 1 \bigr) + r Q_t\bigl( r x+ a, 1 \bigr) \ .
&$y\in (\Lambda_2+ b,1]$ \cr 
} \label{4.4c}
\end{equation}

In terms of $Q_t(x,y)$, the expectation value of 
a dynamical variable $\varphi(x,y)$
is expressed as

\begin{equation}
\int_{[0,1]^2} \varphi(x,y) \rho_{t}(x,y) dx dy = \int_{[0,1]^2} \varphi(x,y) \
dx d_y Q_{t}(x,y)  \ , \label{4.5}
\end{equation}

\noindent
where $d_y$ stands for the Riemann-Stieltjes integral of $Q_t$ with respect to
$y$ \cite{42}.

To solve Eqs. \ (\ref{4.4a}), (\ref{4.4b}) and (\ref{4.4c}), we first
investigate the equation of motion of
$Q_t(x,1)$:

\begin{equation}
Q_{t+1}(x, 1) =l Q_t\bigl( l x, 1 \bigr) + r Q_t\bigl( r x+ a , 1 \bigr) \ .
\label{4.6}
\end{equation}

\noindent
We observe that, in terms of a function $H(x)$ defined by a de Rham equation
(\ref{4.7}), one has

\begin{eqnarray}
 \int_0^1 dH(x') Q_t(x',1) =
\nu \int_0^1 &dH(x')& Q_{t-1}(x',1)\cdots = \nu^t \int_0^1 dH(x') Q_0(x',1)
\ , \label{4.8a} \\
Q_t(x,1)-\int_0^1 dH(x') Q_t(x',1) &=& \int_0^1 dx' \biggl\{ H(x') -
\theta(x'-x)
\biggr\} \partial_{x'}Q_t(x',1) \ , \label{4.8b} 
\end{eqnarray}

\noindent
where $\nu = l+r (\le 1)$ and $\lambda'=\max(l, \ r) (< 1)$.
On the other hand, Eq. (\ref{4.6}) leads to an inequality

$$
{ \Norm{\partial_x Q_t(\cdot,1)} } \le \lambda' 
{ \Norm{\partial_x Q_{t-1}(\cdot,1)}
} \le \cdots \le \lambda'^t { \Norm{\partial_x Q_0(\cdot,1)} } \ , 
$$

\noindent
with ${ \Norm{\partial_x Q_t(\cdot,1)} } \equiv \sup_{x\in[0,1]}|\partial_x
Q_t(x,1)|$. Then, one obtains from Eqs. (\ref{4.8a}) and (\ref{4.8b})

$$
Q_t(x,1)= \nu^t \int_0^1 dH(x') Q_0(x',1) + {\rm O}(\lambda'^t) \ ,
$$

\noindent
and, thus, 

\begin{eqnarray}
R_t(x,y) &=& \nu^t \int_0^1 dH(x') Q_0(x',1) g^{(1)}(y) +{\rm O}(\lambda'^t)
\ , \label{4.9a} \\
g^{(1)}(y) &=& \cases{ 0 \ ,
&$y\in [0,\Lambda_1]$ \cr 
l \ ,
&$y\in (\Lambda_1,\Lambda_2+b]$ \cr  
l+r \ .
&$y\in (\Lambda_2+b,1]$ \cr 
} \label{4.9b} 
\end{eqnarray}

Since $\nu =l+r > \max(l, \ r)=\lambda'$, the lemma of Sec. III can be applied to
the evolution equations (\ref{4.4a}), (\ref{4.4b}) and (\ref{4.4c}) and one
obtains

\begin{equation}
Q_t(x,y) = \nu^t \int_0^1 dH(x') Q_0(x',1) G(y) + {\rm O}(t\lambda'^t) \ ,
\label{4.10a}
\end{equation}

\noindent
where $G(y)$ is the solution of a de Rham equation  (\ref{4.10b}).
The desired result (\ref{4.2}) immediately follows from Eq. (\ref{4.10a}).

\section{ON AN INVARIANT MEASURE $\mu_{ \lcIN }$ FOR A BAKER MAP \\ WITH
ESCAPE}

In this appendix, we show that the invariant measure $\mu_{\rm in}$ on the
fractal repeller considered in Sec. IV is a Gibbs measure.

To show that, we first observe that the image $\Psi^m [0,1]^2$ of the unit
square by the map $\Psi^m$ consists of $2^m$ horizontal strips, which will be
referred to as $H_1^{[m]}, H^{[m]}_2, \cdots H^{[m]}_{2^m}$; and that the
pre-image of
$\Psi^{-n} [0,1]^2$ of the unit square by the map $\Psi^n$ consists of $2^n$
vertical strips, which will be referred to as $V^{[n]}_1, V^{[n]}_2, \cdots
V^{[n]}_{2^n}$. Then, boxes $H^{[m]}_i \cap V^{[n]}_j$ generated by the overlap
procedure provide a generating partition of the repeller. As easily seen, for
each box
$H^{[m]}_i \cap V^{[n]}_j$, a stretching factor for a time interval $[-m,n-1]$

\begin{equation}
u_{ij}(n,m) \equiv \sum_{t=-m}^{n-1}
\lambda_x\biggl(\Psi^t(x,y)\biggr) \ , \label{F.1}
\end{equation}

\noindent
does not depend on the initial point $(x,y) \in H^{[m]}_i \cap V^{[n]}_j$, where
$\lambda_x(x,y)$ is  the local expanding rate defined by
$\lambda_x(x,y)=- \ln l$ (if $x\in [0,l]$) and $\lambda_x(x,y)=- \ln
r$ (if $x\in [a,a+r]$).
Note that, when $(x,y)\in H^{[m]}_i \cap V^{[n]}_j$, the pre-image
$\Psi^{-k}(x,y)$ is unique for $k=1,2,\cdots m$.  We show that the
$\mu_{\rm in}$-measure of the box $H^{[m]}_i \cap V^{[n]}_j$ is given by

\begin{equation}
\mu_{\rm in}\bigl(H^{[m]}_i \cap V^{[n]}_j\bigr) = {e^{- u_{ij}(n,m)} \over
\sum_{i,j} e^{- u_{ij}(n,m)} } \ , \label{F.2}
\end{equation}

\noindent
which implies that the measure $\mu_{\rm in}$ is a Gibbs measure
\cite{1,4,5,20b}. 
Note that, since the numerator of Eq. (\ref{F.2}) is a product $l^s
r^{n+m-s}$ of $(n+m)$ factors ($s=0,1,2,\cdots n+m$) and there are
$(n+m)!/\{s!(n+m-s)!\}$ boxes with this value, the sum of the
numerators, or the normalization factor, is

$$
\sum_{i,j} e^{- u_{ij}(n,m)} = \sum_{s=0}^{n+m}
{(n+m)!\over s!(n+m-s)! }l^s r^{n+m-s}=(l+r)^{n+m}=e^{-(n+m) \kappa } \ ,
$$

\noindent
where $\kappa = - \ln(l+r)$ is the escape rate \cite{20b}.

Before proving Eq. (\ref{F.2}), we verify it for a simple case $n=2$ and $m=1$. 
In this case, we have $V_1^{[2]}=[0,l^2]\times [0,1]$, $V^{[2]}_2=[a,rl+a]\times
[0,1]$, 
$V^{[2]}_3=[la,lr+la]\times [0,1]$,  $V^{[2]}_4=[ra+a,r^2+ra+a]\times [0,1]$;
$H^{[1]}_1=[0,1]\times [0,\Lambda_1]$ and $H^{[1]}_2=[0,1]\times [b,\Lambda_2 +
b]$. As an example, we consider a box $H^{[1]}_2 \cap V^{[2]}_3=[la,lr+la]\times
[b,\Lambda_2 + b]$.  For any point $(x,y) \in H^{[1]}_2 \cap V^{[2]}_3$, we have
$0\le x \le l$ and $\Psi(x,y), \Psi^{-1}(x,y)  \in [a,a+r]\times [0,1]$ and,
thus, 

$$
u_{2,3}(2,1) \equiv \sum_{t=-1}^{1}
\lambda_x\biggl(\Psi^t(x,y)\biggr)
=  - 2 \ln r - \ln l  \ .
$$

\noindent
On the other hand, from the functional equations of $H(x)$ and $G(x)$, we have

\begin{eqnarray}
\mu_{\rm in}\bigl(H^{[1]}_2 \cap V^{[2]}_3\bigr) &=& \{ H(lr+la)-H(la) \} \{
G(\Lambda_2 + b)-G(b)\} \nonumber \\
&=& {l\over l+r}{r\over l+r}\{ H(1)-H(0) \}{r\over l+r}\{ G(1)-G(0)\} \nonumber
\\ 
&=& {lr^2\over (l+r)^3}={\exp(- u_{2,3}(2,1)) \over (l+r)^3} \ , \nonumber
\end{eqnarray}

\noindent
which is (\ref{F.2}). 

Now we go to the proof of Eq. (\ref{F.2}). Since one can write
$H^{[m]}_i=[0,1]\times [\alpha_i^{[m]},\beta_i^{[m]}]$ and
$V^{[n]}_j=[\gamma_j^{[n]},\delta_j^{[n]}]
\times [0,1]$, and thus,

\begin{eqnarray}
\mu_{\rm in}\bigl(H^{[m]}_i \cap V^{[n]}_j\bigr)&=&\{G(\beta_i^{[m]})-
G(\alpha_i^{[m]})\}\{H(\delta_j^{[n]})-H(\gamma_j^{[n]})\} \nonumber \\
&=& \mu_{\rm in}\bigl(H^{[m]}_i\bigr) \mu_{\rm in}\bigl(V^{[n]}_j\bigr) \ ,
\nonumber 
\end{eqnarray}

\noindent
it is enough to show

\begin{equation}
\mu_{\rm in}\bigl( V^{[n]}_j\bigr) = {e^{- u_{j}(n)} \over (l+r)^n } \ ,
\label{F.3}
\end{equation}

\noindent
and

\begin{equation}
\mu_{\rm in}\bigl(H^{[m]}_i \bigr) = {e^{- u_{i}(m)} \over
(l+r)^m } \ , \label{F.4}
\end{equation}

\noindent
where $u_{j}(n)=\sum_{t=0}^{n-1} \lambda_x\biggl(\Psi^t(x,y)\biggr)$ and
$u_{i}(m)=\sum_{t=-m}^{-1} \lambda_x\biggl(\Psi^t(x,y)\biggr)$ are stretching
factors for a vertical strip $V^{[n]}_j$ and a horizontal strip $H^{[m]}_i$,
respectively. As before, the stretching factors are constant on each strip.
The relations (\ref{F.3}) and (\ref{F.4}) are proved by induction with respect
to $n$ and $m$, with the aid of the functional equations for $H(x)$ and $G(y)$,
respectively. Since the proofs of the two relations are similar, we show
(\ref{F.3}) only. 

It is easy to see that (\ref{F.3}) holds for $n=1$.
Now we suppose that Eq. (\ref{F.3}) is valid for $n$. As easily seen, a
vertical strip $V^{[n+1]}_{j'}$ is expressed by some vertical strip $V^{[n]}_j$
as

$$
V^{[n+1]}_{j'} = R_\sigma \cap \Psi^{-1} V^{[n]}_j
$$

\noindent
where $\sigma=0$ or 1 with $R_0=[0,l]\times [0,1]$ and $R_1=[a,a+r]\times
[0,1]$. In terms of $\gamma_j^{[n]}$ and $\delta_j^{[n]}$, we have

$$
V^{[n+1]}_{j'} = \cases{[l \gamma_j^{[n]}, l \delta_j^{[n]}] \times [0,1] \ ,
&$\sigma=0$ \cr
[r\gamma_j^{[n]}+a,r\delta_j^{[n]}+a] \times [0,1] \ . &$\sigma=1$ \cr}
$$

\noindent
When $\sigma=0$, from the functional equation for $H(x)$, one obtains

\begin{eqnarray}
\mu_{\rm in}(V^{[n+1]}_{j'}) &=& H(l \delta_j^{[n]})-H(l \gamma_j^{[n]})
= {l\over l+r} \{H( \delta_j^{[n]})-H( \gamma_j^{[n]})\} \nonumber \\
&=& {l\over l+r} \mu_{\rm in}(V^{[n]}_{j}) 
= {\exp\Bigl({- u_{j}(n)+ \ln l}\Bigr) \over (l+r)^{n+1} } \ .
\end{eqnarray}

\noindent
Then, for $(x,y)\in V^{[n+1]}_{j'}$, one has $(x,y)\in R_0$ and $\Psi(x,y)\in
V^{[n]}_j$. Therefore, $\lambda_x(x,y)=- \ln l$ and

\begin{equation}
u_j(n)-\ln l = \sum_{t=0}^{n-1} \lambda_x\biggl(\Psi^t\circ \Psi(x,y)\biggr) 
+ \lambda_x(x,y) = \sum_{t=0}^{n} \lambda_x\biggl(\Psi^t(x,y)\biggr)
=u_{j'}(n+1) \ . \label{F.F}
\end{equation}

\noindent
Similarly, one can verify Eq. (\ref{F.F}) when $\sigma=1$. Hence, 

$$
\mu_{\rm in}(V^{[n+1]}_{j'}) = {\exp\Bigl(-u_{j'}(n+1)\Bigr) \over
(l+r)^{n+1} } \ ,
$$

\noindent
or Eq. (\ref{F.3}) holds for $n+1$ and, by induction, it is valid for
all positive integer $n$.

\section{THE EVOLUTION EQUATION OF MEASURES FOR \\ A REACTION MODEL}

In this Appendix, we write down the evolution equation of the partially
integrated distribution function $Q_t(\alpha:x,y)=\int_0^y dy'
\rho_t(\alpha:x,y')$ ($\alpha=R, I$, or $P$) for a chemical reaction model
introduced in Sec. V. 
The density function $\rho_t(\alpha:x,y)$ ($\alpha=R, I$, or $P$) at time $t$
is given by

$$
\rho_t(\alpha:x,y) = \sum_{\beta=R,I,P} \int_{[0,1]^2} dx dy
\delta\left((\alpha:x,y)-\Psi'^t(\beta:x',y')\right) \rho_0(\beta:x',y') \ ,
$$

\noindent
where $\rho_0$ is the density function of the initial measure, $\Psi'$ is the map
introduced in Sec. V (cf. Eqs. (\ref{Imap}), (\ref{Rmap}), and (\ref{Pmap}) ) 
and the delta function $\delta\left((\alpha:x,y)-(\beta:x',y')\right)$ stands for
the product $\delta_{\alpha,\beta} \delta(x-x') \delta(y-y')$.

By integrating the Frobenius-Perron equation for the density 
$\rho_t(\alpha:x,y)$, one obtains the evolution equation for $Q_t$ :

\begin{equation}
Q_{t+1}(\alpha:x,y) = {\bar {\cal F}}Q_t(\alpha:x,y) + {\bar R}_t(\alpha:x,y)
\ , \label{EvEQ2}
\end{equation}

\noindent
where $\alpha=R, I$, or $P$, a linear contraction mapping ${\bar {\cal F}}$ is
defined by

\FL
\begin{eqnarray}
{\bar {\cal F}}Q_t(I:x,y) &=& \cases{ l Q_t\left(I: lx,{\displaystyle y \over
\displaystyle \Lambda_1}
\right) \ , &$y\in (0,\Lambda_1]$
\cr 
(b-\Lambda_1) Q_t \left(R: (b-\Lambda_1) x, {\displaystyle y-\Lambda_1 \over
\displaystyle b-\Lambda_1} \right) \ , &$y\in (\Lambda_1,b]$
\cr 
r Q_t \left(I: rx+a, {\displaystyle y-b \over \displaystyle \Lambda_2}\right)
\ , &$y\in (b,\Lambda_2+b]$
\cr
(1-b-\Lambda_2) Q_t\left(P: (1-b-\Lambda_2) x, {\displaystyle y-b-\Lambda_2
\over \displaystyle 1-b-\Lambda_2} \right) \ , &$y\in (\Lambda_2+b,1]$ \cr}
\label{Imeas}  \\ \nonumber \\ \nonumber \\
{\bar {\cal F}}Q_t(R\col x,y) &=& \cases{ (a-l) Q_t\left(I\col (a-l)x+l,
{\displaystyle y \over \displaystyle b-\Lambda_1} \right)
\ , &$y\in (0,b-\Lambda_1]$ \cr \cr
(1-b+\Lambda_1) Q_t\left(R\col (1-b+\Lambda_1) x+b-\Lambda_1 , {\displaystyle
y-b+\Lambda_1 \over \displaystyle 1-b+\Lambda_1}\right) \ , &$y\in
(b-\Lambda_1,1]$ \cr}
\label{Rmeas}  \\ \nonumber \\ \nonumber \\
{\bar {\cal F}}Q_t(P\col x,y) &=& \cases{ (1\hik a\hik r)
Q_t\left(I\col (1\hik a\hik r)x\tas a\tas r, {\displaystyle y
\over \displaystyle 1-b-\Lambda_2}\right) \ , &$y\in (0,1\hik b\hik \Lambda_2]$
\cr
\cr (b+\Lambda_2) Q_t\left(P\col (b\tas \Lambda_2) x\tas 1\hik b\hik \Lambda_2 ,
{\displaystyle y-1+b+\Lambda_2 \over \displaystyle
b+\Lambda_2} \right)
\ , &$y\in (1\hik b\hik \Lambda_2,1]$ \cr}
\label{Pmeas}
\end{eqnarray}

\noindent
and  ${\bar R}_t(\alpha:x,y)$ is a functional of $Q_t(\alpha: x, 1)$ :

\FL
\begin{eqnarray}
{\bar R}_t(I:x,y) &=& \cases{ 0 \ , &$y\in (0,\Lambda_1]$
\cr 
l Q_t \left(I: l x, 1 \right) \ , &$y\in (\Lambda_1,b]$
\cr 
l Q_t \left(I: l x, 1 \right) + (b-\Lambda_1) 
Q_t \left(R: (b-\Lambda_1) x, 1 \right)
\ , &$y\in (b,\Lambda_2+b]$
\cr
l Q_t \left(I: l x, 1 \right) + (b-\Lambda_1) 
Q_t \left(R: (b-\Lambda_1) x, 1 \right) \cr
~~~~~~~~~~~~~~~~~~~~~~~~~~~~~~~~~+r Q_t\left(I: r x+a, 1 \right)
\ , &$y\in (\Lambda_2+b,1]$ \cr}
\label{Iresi}  \\ \nonumber \\ \nonumber \\
{\bar R}_t(R:x,y) &=& \cases{ 0
\ , &~~~~~~~~~~~$y\in (0,b-\Lambda_1]$ \cr
(a-l) Q_t\left(I: (a-l)x+l, 1 \right) \ , &~~~~~~~~~~~$y\in
(b-\Lambda_1,1]$ \cr}
\label{Rresi}  \\ \nonumber \\ \nonumber \\
{\bar R}_t(P:x,y) &=& \cases{ 0 \ , &$y\in (0,1-b-\Lambda_2]$ \cr
(1-a-r) Q_t\left(I:(1-a-r)x+a+r, 1 \right)
\ . &$y\in (1-b-\Lambda_2,1]$ \cr}
\label{Presi}
\end{eqnarray}

\noindent
These are the desired results. Note that the contraction constant $\bar \lambda$
of the mapping ${\bar {\cal F}}$ is given by

$$
{\bar \lambda} = \max\biggl(l,r,a-l,1-a-r,b-\Lambda_1,b+\Lambda_2,1-b+\Lambda_1,
1-b-\Lambda_2\biggr) (<1) \ .
$$

\newpage

\begin{figure}
\caption{Weierstrass function for $a=2/3$ and $b=2$.~~~~~~~~~~~~~~
~~~~~~~~~~~~~~~~~~~~~~~~~~~~~~~~~~~~~~~~~~~~~~~~~~~~~~~~~~~}
\label{fig1}
\end{figure}

\begin{figure}
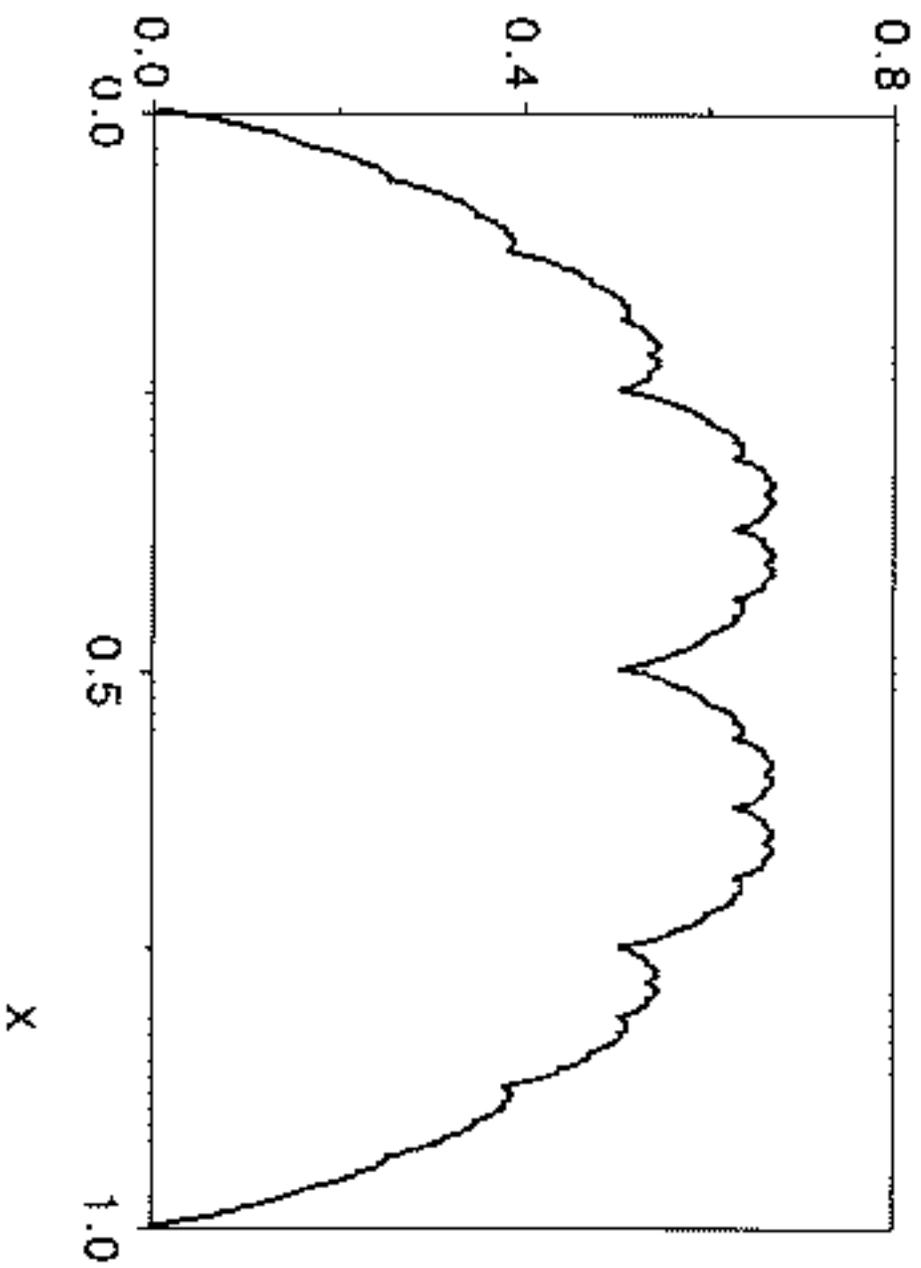

\caption{Takagi function.~~~~~~~~~~~~~~~~~~~~~~~~~~~~~
~~~~~~~~~~~~~~~~~~~~~~~~~~~~~~~~~~~~~~~~~~~~~~~~~~~~~~~~}
\label{fig2}
\end{figure}

\begin{figure}
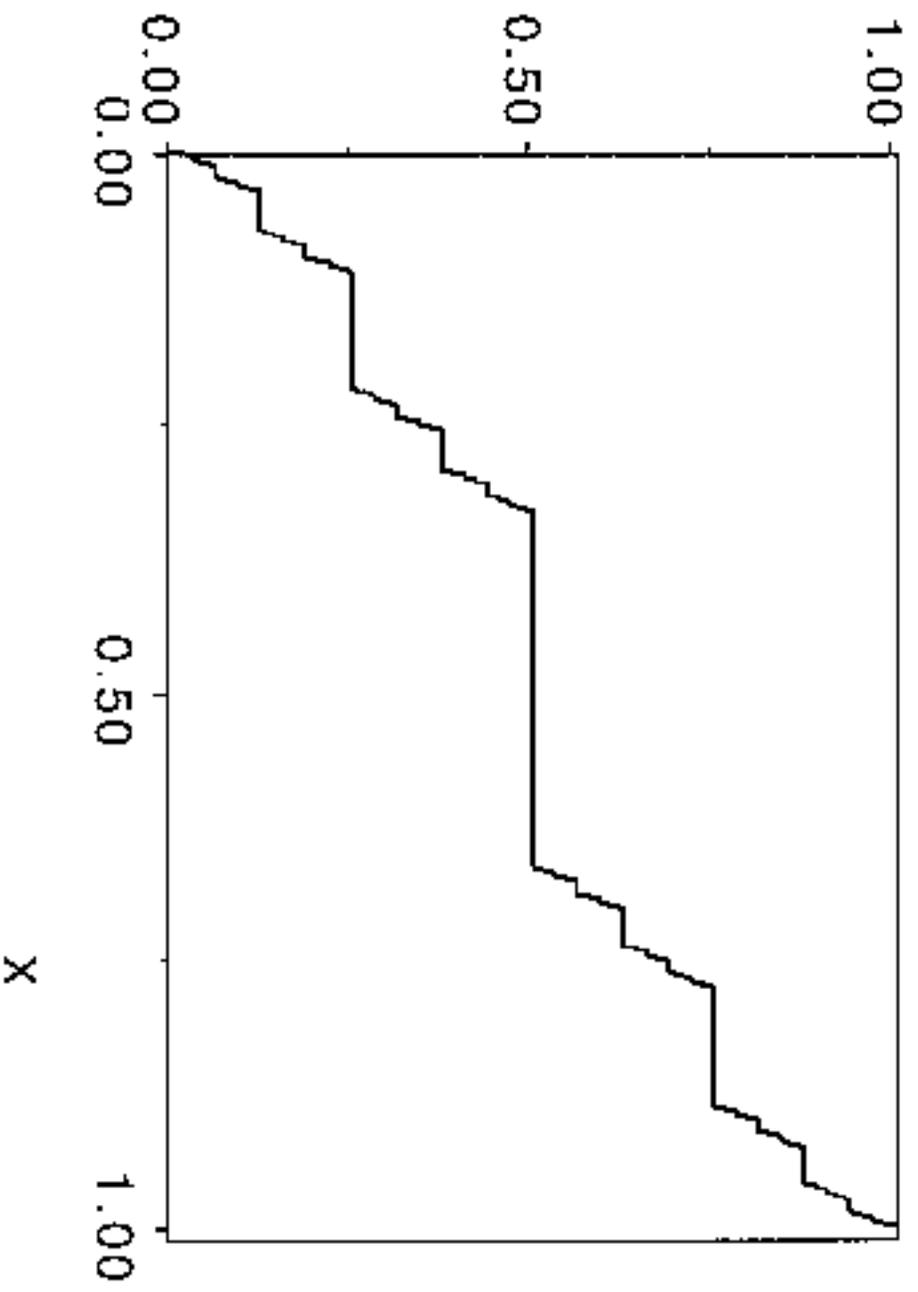

\caption{Cantor function.~~~~~~~~~~~~~~~~~~~~~~~~~~~~~
~~~~~~~~~~~~~~~~~~~~~~~~~~~~~~~~~~~~~~~~~~~~~~~~~~~~~~~~}
\label{fig3}
\end{figure}

\begin{figure}
\caption{Lebesgue's singular function for $\alpha=0.75$.
~~~~~~~~~~~~~~~~~~~~~~~~~~~~~~~~~~~~~~~~~~~~~~~~~~~~}
\label{fig4}
\end{figure}

\begin{figure}
\caption{Schematic representation of the non-conservative reversible Baker 
map. The shaded rectangle is expanded and the rest is contracted.}
\label{fig5}
\end{figure}

\begin{figure}
\caption{Partially integrated distribution of the 
physical measure $\mu_{\rm ph}$ along $y$ for the 
non-conservative reversible map with $l=0.3$.}
\label{fig6}
\end{figure}

\begin{figure}
\caption{Schematic representation of the Baker map with a Cantor-like invariant
set. The points in the black rectangle are removed. The  other shaded rectangles
are mapped onto the corresponding ones.}
\label{fig7}
\end{figure}

\begin{figure}
\caption{Partially integrated distribution of the 
physical measure $\mu_{\rm ph}$ along $x$ for the Baker map with a Cantor-like
invariant set. The parameters are
$\Lambda_1=0.4$, $\Lambda_2=0.3$, $b=0.5$ and $l/r=0.8$.}
\label{fig8}
\end{figure}

\begin{figure}
\caption{Schematic representation of the Baker-type map describing a
chemical reaction process $R \leftrightarrow I \leftrightarrow P$.
Three unit squares describe the dynamical states of the reactant, intermediate
and product, respectively.  The shaded rectangles are mapped onto the
corresponding ones.}
\label{figFL}
\end{figure}

\begin{figure}
\caption{Partially integrated distribution of the 
stationary measure $\mu_{\rm fl}$ along $x$ for the 
reaction model under a flux boundary condition. The parameters are chosen as
$\Lambda_1=0.4$, $\Lambda_2=0.3$, $b=0.5$, $l=0.4$, $r=0.5$, $R_{\rm ex}=0$,
and $P_{\rm ex}=(1-l-r)/(1-b-\Lambda_2)$.}
\label{fig9}
\end{figure}


\begin{references}
\bibitem{42} This means that, for a partition $0=y_0<y_1<\cdots<y_n=1$ of
the interval [0,1] and $y_{k-1}\le \xi_k \le y_k$, 
$$
\int_0^1 \varphi(x,y) d_yP_t(x,y) = \lim_{{\scriptscriptstyle
\max(y_k-y_{k-1})}\to 0} \ \ \ \sum_{k=1}^n  \varphi(x,\xi_k)
\{P_t(x,y_k)-P_t(x,y_{k-1})\} \ . 
$$
\end{references}
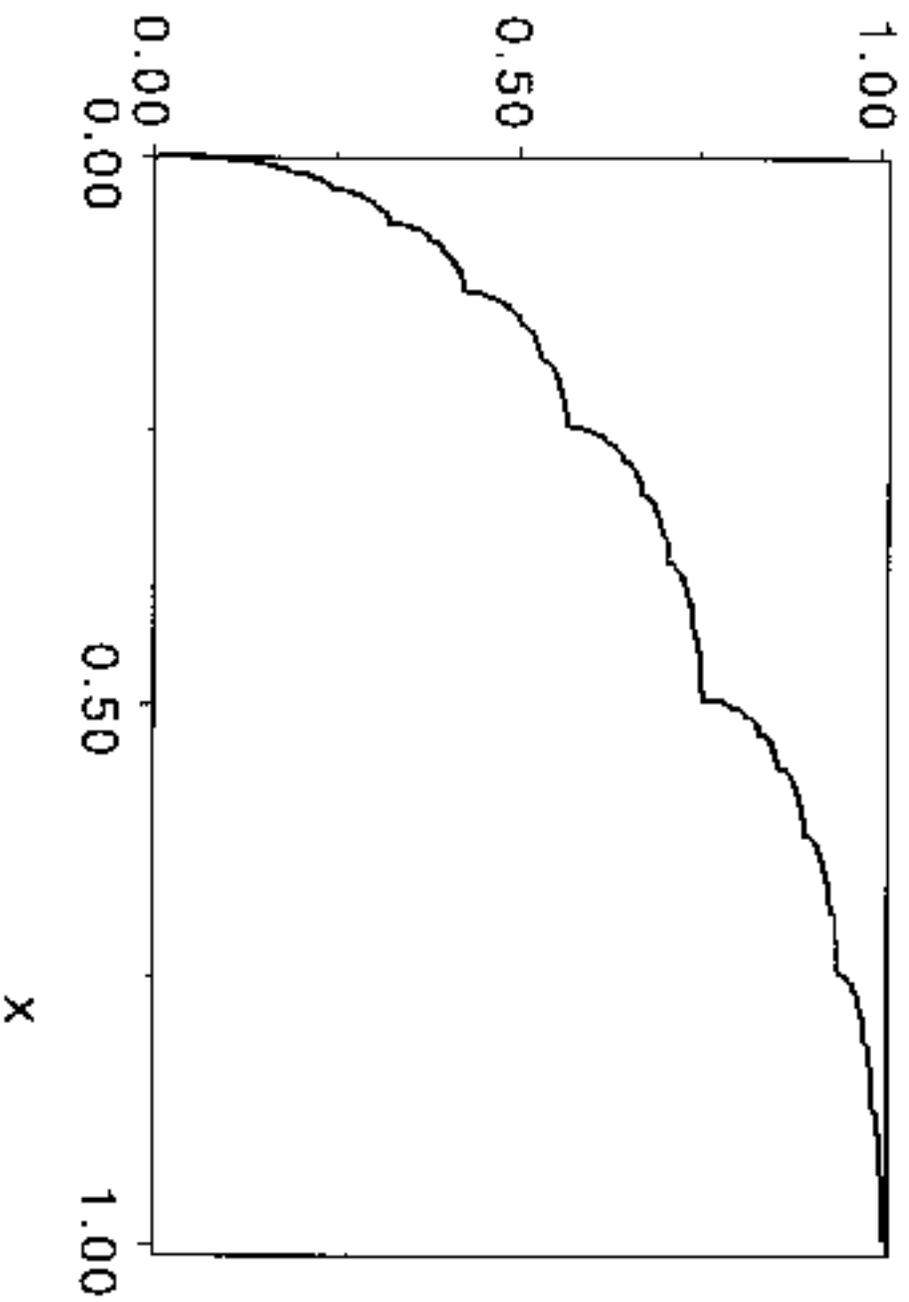
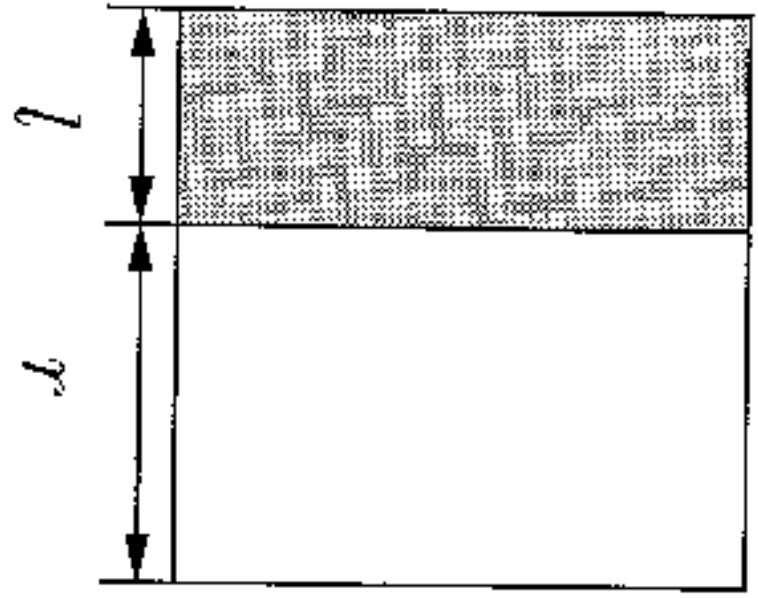
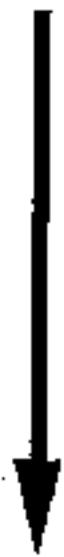
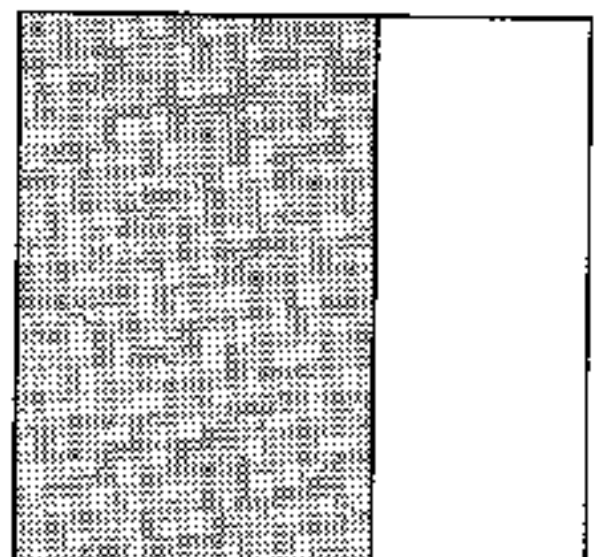
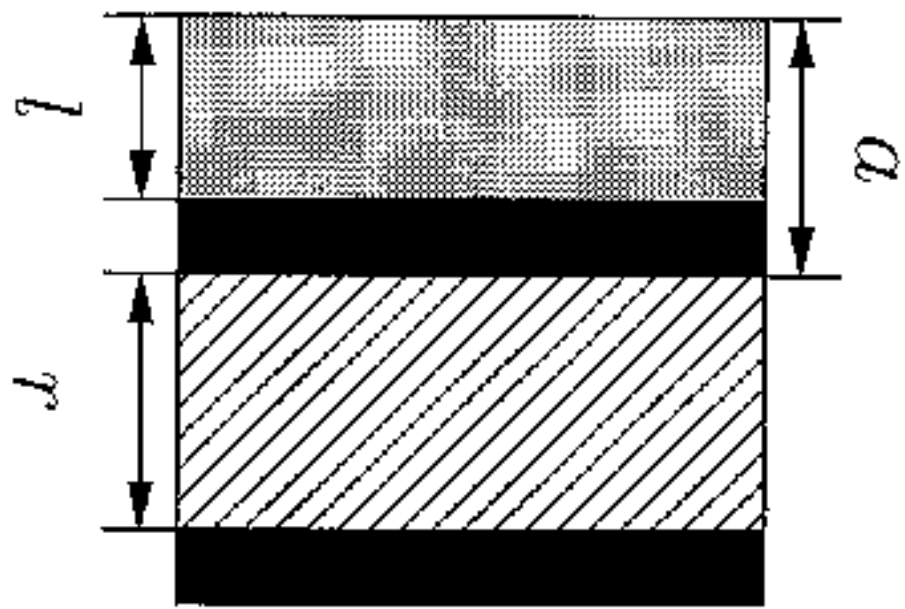
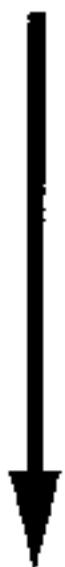
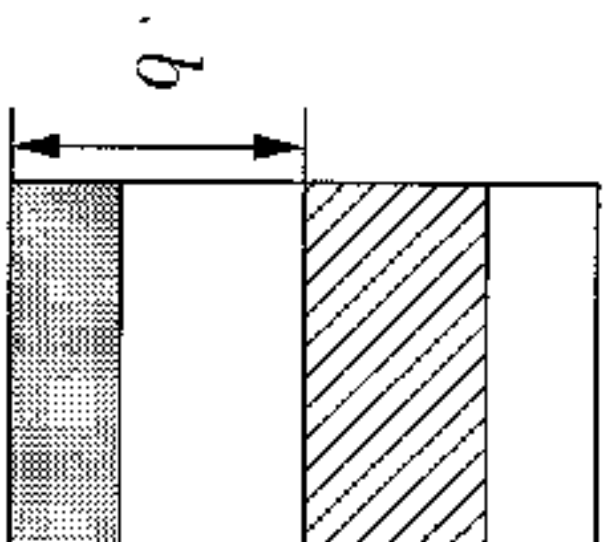
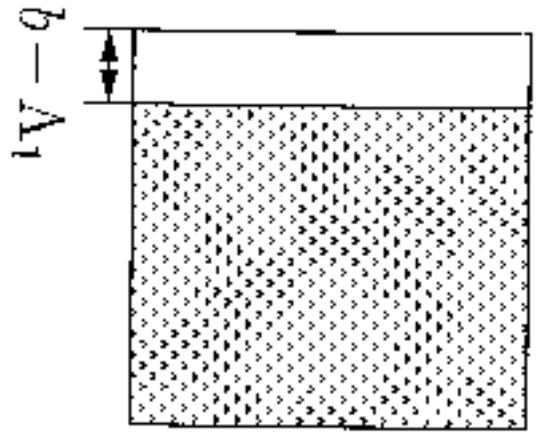
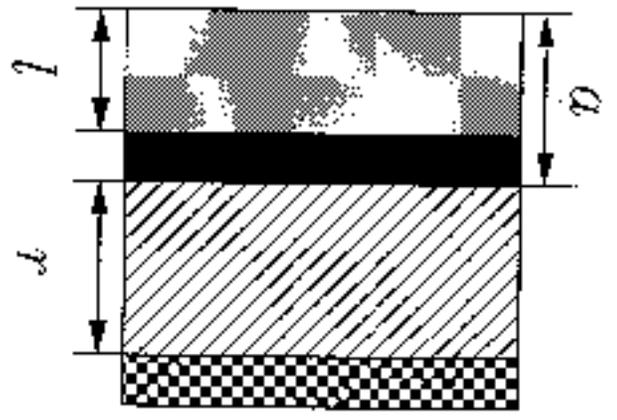
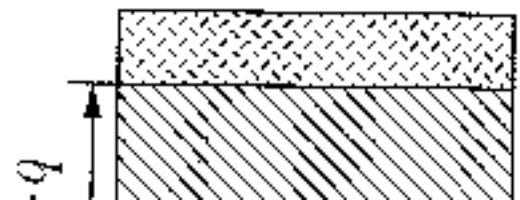
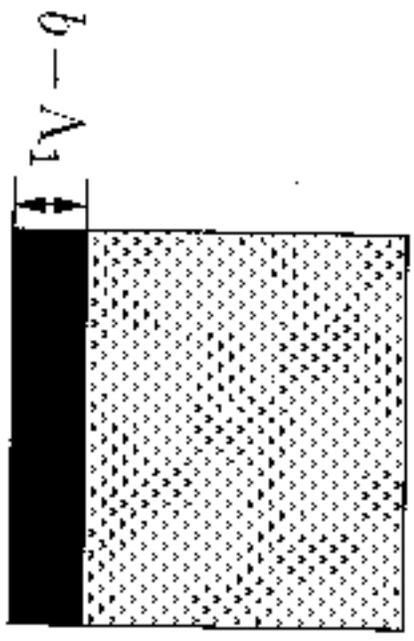
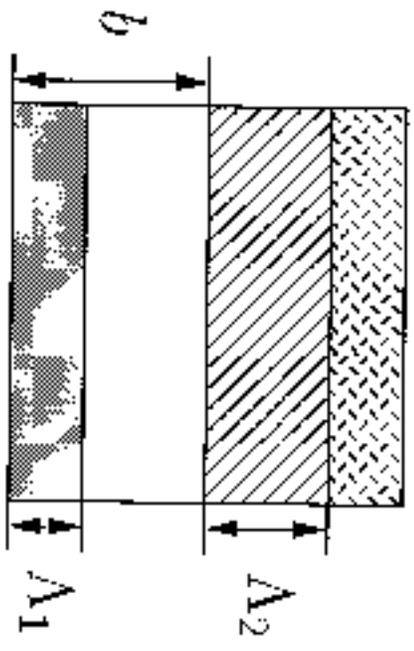
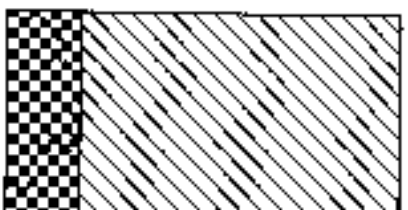
\end{document}